\RequirePackage[safe,debrief]{silence}  
     \ActivateFilters[acmart-family,todonotes-family,acronym-in-section-titles-family,acmart-family]  

\documentclass[sigconf,natbib=false,authorversion]{acmart} 

\copyrightyear{2026}
\acmYear{2026}
\setcopyright{cc}
\setcctype[4.0]{by}
\acmConference[SAC '26]{The 41st ACM/SIGAPP Symposium on Applied Computing}{March 23--27, 2026}{Thessaloniki, Greece}
\acmBooktitle{The 41st ACM/SIGAPP Symposium on Applied Computing (SAC '26), March 23--27, 2026, Thessaloniki, Greece}
\acmDOI{10.1145/3748522.3779717}
\acmISBN{979-8-4007-2294-3/2026/03}

\RequirePackage[
  datamodel=acmdatamodel,
  style=acmnumeric,
  backend=biber,
  ]{biblatex}

\graphicspath{{./figs/}}

\usepackage[T1]{fontenc}
\usepackage{acronym}
\usepackage{listings}
\usepackage{amsmath,amsfonts} 
\usepackage{fontawesome}

\usepackage{csquotes}
\usepackage{ragged2e}
\usepackage{array}  

\begin{document}

\title{A Practical Solution to Systematically Monitor Inconsistencies in SBOM-based Vulnerability Scanners}
\renewcommand{\shorttitle}{Exploring Inconsistencies in SBOM-Based Vulnerability Scanners}

\author{Martin Rosso}
\email{martin.rosso@unipd.it}
\affiliation{%
  \institution{University of Padua}
  \city{Padua}
  \country{Italy}
}

\author{Muhammad Asad Jahangir Jaffar}  
\email{muhammad.jaffar@kineton.it}
\affiliation{%
  \institution{Kineton S.r.l.}
  \city{Naples}
  \country{Italy}
}

\author{Alessandro Brighente}
\email{alessandro.brighente@unipd.it}
\affiliation{%
  \institution{University of Padua}
  \city{Padua}
  \country{Italy}
}

\author{Mauro Conti}
\email{mauro.conti@unipd.it}
\affiliation{%
  \institution{University of Padua}
  \institution{Örebro University}
  \city{Padua}
  \country{Italy}
  \city{Örebro}
  \country{Sweden}
}

\renewcommand{\shortauthors}{M. Rosso et al.}




\newcommand{\appendixref}[1]{\hyperref[#1]{Appendix~\ref*{#1}}}

\newacro{cisa}[CISA]{U.S. Cybersecurity and Infrastructure Security Agency}
\newacro{nist}[NIST]{U.S. National Institute of Standards and Technology}
\newacro{nvd}[NVD]{National Vulnerability Database}
\newacro{BOM}[BOM]{Bill of Materials}
\newacro{SBOM}[SBOM]{Software Bill of Materials}
\newacro{ssc}[SSC]{Software Supply Chain}
\newacro{vex}[VEX]{Vulnerability Exploitability eXchange}
\newacro{purl}[purl]{package URL}
\newacro{CPE}[CPE]{Common Platform Enumeration}
\newacro{CVE}[CVE]{Common Vulnerabilities and Exposures}
\newacro{osv}[OSV]{Google Open Source Vulnerabilities}
\newacro{oss}[OSS Index]{Sonatype OSS Index}
\newacro{ghsa}[GHSA]{GitHub Security Advisory}

\newacro{OWASP}[OWASP]{Open Worldwide Application Security Project}

\newacro{sca}[SCA]{Software Composition Analysis}
\newacro{svs}[SVS]{SBOM-based Vulnerability Scanning}
\newacro{uri}[URI]{Uniform Resource Identifier}

\newacro{eu}[EU]{European Union}
\newacro{NTIA}[NTIA]{U.S. National Telecommunications and Information Administration}
\newacro{SPDX}[SPDX]{System Package Data Exchange}

\newacro{SWID}[SWID]{Software Identification Tags}
\newacro{SWHID}[SWHID]{Software Hash Identifier}
\newacro{gitoid}[gitoid]{Git Object ID}

\newacro{SOC}[SOC]{(Cyber) Security Operations Center}

\newcommand{\svs}[0]{\ac{svs}}
\newcommand{\svstool}[0]{\svs{}-tool}
\newcommand{\svsaas}[0]{\svs{}-as-a-Service}
\newcommand{\cpe}[1]{{\texttt{\textlangle #1\textrangle}}}  
\newcommand{\purl}[1]{\texttt{\textlangle #1\textrangle}}  
\newacro{svstts}[SVS-TEST]{\svstool{} Testing Solution}
\newcommand{\svstts}[0]{\acs{svstts}}
\newcommand{\experimentaleval}[0]{case study}  

\newcommand{\wildsbomdataset}[0]{\emph{Wild SBOM} dataset}





\newcommand{\bomber}[0]{Bomber}
\newcommand{\depscan}[0]{dep-scan}
\newcommand{\dtrack}[0]{Dependency-Track}
\newcommand{\grype}[0]{Grype}
\newcommand{\snyk}[0]{snyk-cli}
\newcommand{\cast}[0]{CAST-Highlights}
\newcommand{\fossa}[0]{FOSSA}
\newcommand{\vulert}[0]{Vulert}

\newcommand{\cve}[1]{\href{https://www.cve.org/CVERecord?id=#1}{#1}}  
\newcommand{\iso}[1]{#1}  
\newcommand{\rfc}[1]{\href{https://www.rfc-editor.org/rfc/rfc#1}{RFC~#1}}
\newcommand{\ecma}[1]{\href{https://ecma-international.org/publications-and-standards/standards/ECMA-424/#1}{#1}}

\newcommand{\doi}[1]{\href{https://www.doi.org/#1}{doi:#1}}
\newcommand{\artifactrepo}[0]{artifact repository}
\newcommand{\artifactrepolink}[0]{\doi{10.5281/zenodo.17921254}}

\newcommand{\requirement}[1]{\texttt{R#1}}
\newcommand{\phase}[1]{phase~#1} 
\newcommand{\scenario}[1]{Scenario~#1}
\newcommand{\testcase}[1]{\texttt{#1}}

\newcommand{\Example}[0]{\smallskip{}\noindent{}\textit{Example:}}  

\newcommand{\strongemph}[1]{\textbf{\emph{#1}}}

\newcommand{\numsvstools}[0]{7}  
\newcommand{\numtestcases}[0]{16}  
\newcommand{\numtestscenarios}[0]{8}

\newcommand{\tick}[0]{\ding{52}}
\newcommand{\cross}[0]{\ding{54}}

\newcommand{\figurecomment}[1]{ 
                                \begin{minipage}{0.98\linewidth}
                                    \scriptsize #1
                                \end{minipage}
                                }
                                
\newcommand{\tablecomment}[1]{ \figurecomment{#1} }

\newcommand{\pseudoparagraph}[1]{\smallskip\noindent\textit{#1}}
\newcommand{\boldpseudoparagraph}[1]{\smallskip\noindent\textbf{#1.}}

\newcommand{\testcasetitle}[2]{\noindent\textbf{Test Case \testcase{#1}} (#2):}

\newcommand{\code}[1]{\texttt{#1}}
\renewcommand{\code}[1]{\lstinline[keywordstyle=\ttfamily]`#1`}
\newcommand{\component}[1]{\texttt{#1}}
\newcommand{\componentp}[2]{\texttt{({#1})\,#2}}
\newcommand{\version}[1]{\texttt{#1}}

\newcommand{\hint}[1]{\smallskip\begin{minipage}[c]{0.9\linewidth}\hspace{-1.5em}\ding{43} #1 \end{minipage}}  

\begin{abstract}
  Software Bill of Materials (SBOM) provides new opportunities for automated vulnerability identification in software products.
While the industry is adopting SBOM-based Vulnerability Scanning (SVS) to identify vulnerabilities, we increasingly observe inconsistencies and unexpected behavior, that result in false negatives and silent failures.
In this work, we present the background necessary to understand the underlying complexity of SVS and introduce SVS-TEST, a method and tool to analyze the capability, maturity, and failure conditions of SVS-tools in real-world scenarios.
We showcase the utility of SVS-TEST in a case study evaluating seven real-world SVS-tools using 16 precisely crafted SBOMs and their respective ground truth.
Our results unveil significant differences in the reliability and error handling of SVS-tools;
multiple SVS-tools silently fail on valid input SBOMs, creating a false sense of security.
We conclude our work by highlighting implications for researchers and practitioners, including how organizations and developers of SVS-tools can utilize SVS-TEST to monitor SVS capability and maturity. 
All results and research artifacts are made publicly available and all findings were disclosed to the SVS-tool developers ahead of time. 


\end{abstract}



\maketitle

\section{Introduction}

Since \citeyear{owasp-top-ten-2013}, the \ac{OWASP} Top~10 project recurrently named the usage of outdated and known to be vulnerable software dependencies among their Top~10 security risks for (web) applications~\cite{owasp-top-ten-2013}.
This is confirmed also in the \citeyear{owasp_risk} Open Source Software Risks: \ac{OWASP} dedicates at least four Top~10 risks to dependency-induced threats~\cite{owasp_risk}.
In response to these growing risks, the United States promoted the use of \ac{SBOM} with the \href{https://www.whitehouse.gov/briefing-room/presidential-actions/2021/05/12/executive-order-on-improving-the-nations-cybersecurity/}{2021 US Executive Order on Improving the Nation's Cybersecurity}. 
While the European \href{http://data.europa.eu/eli/reg/2024/2847/oj}{Cyber Resilience Act~(CRA)} and \href{http://data.europa.eu/eli/dir/2024/2853/oj}{Product Liability Directive~(PLD)} do not strictly mandate the use of \ac{SBOM}, they explicitly confirm the manufacturers' responsibility to deliver secure (software/smart) products and to provide security updates in a timely manner throughout their lifetime -- something that \ac{SBOM} is promising to achieve.
\ac{SBOM} provides a structured way to list software components, and is thus a key technology facilitating automated vulnerability scanning via \acf{svs}.
As a result, \ac{SBOM} and \ac{svs} recently gained (new) traction in academia and industry alike.

To maximize utility of this technology, it is important that researchers and practitioners understand the challenges, complexities, and limitations of \ac{svs}.
In practice, however, we observe significant dissonance between domain experts and practitioners.
On the one hand, the academic literature identified a \enquote{lack of maturity in SBOM tooling} and the need for \enquote{more reliable, user-friendly, standard-conformable, and interoperable enterprise-level SBOM tools}~\cite{sbom-where-we-stand} to mature the technology.
On the other hand, we observe how the industry and practitioners perceive \ac{svs} and \svstool{}s as solution to software security and to comply with legal requirements, but underestimate the limitations of this technology~\cite{linux-foundation-sbom-readiness-survey,onekey}.


Our work directly targets this dissonance.
First, we summarize the \ac{svs} process, highlighting key challenges and limitations and uncovering the hidden complexity of \svstool{}s in an extensive \hyperref[sec:background]{Background} section.
Once a theoretical background is established, we shift our focus to real-world \svstool{}s and how these tools operate in practice and how their limitations and shortcomings may impact an organization's perceived security in practice.
To do so, we develop \svstts{}, a methodology to experimentally test \svstool{}s in realistic edge-case scenarios (i.e., attacks are explicitly not in scope). 
We present the utility of our solution in a \experimentaleval{}.
Evaluating a total of \numsvstools{} established \svstool{}s against a manually created ground truth dataset of \numtestcases{} realistic \ac{SBOM}s, we observe major differences in the way that reliable \svstool{}s operate and what implicit assumptions they make on the input \ac{SBOM}.
Our results show that in practice, \svstool{}s may silently fail to lookup components from the input \ac{SBOM}s, thus creating a false sense of security.
This work has direct impact on \svstool{} users, developers, and the \ac{svs} community as a whole.
We transparently informed \svstool{} developers of our findings resulting in prompt updates of the respective tools.
We transparently make our test cases and results available in our artifact repository.\footnote{All artifacts available at \artifactrepolink{}.}

\section{Background and Related Literature}
\label{sec:background}

In this section, we introduce the concept of \acf{SBOM}, how the industry performs component identification, and present the benefits and limitations of \svstool{}s.

\subsection[Software Bill of Materials (SBOM)]{\acf{SBOM}}
A \acf{BOM} is a structured inventory of all components constituting a product.
While the traditional \ac{BOM} describes a physical product, an \ac{SBOM} focuses on software and digital components.
Typically, an \ac{SBOM} contains metadata about the product itself, how and when the \ac{SBOM} was created, and a list (or tree) of components~\cite{ntia-sbom-minimal-elements} with their transitive relation in a structured machine-readable format.
Two widely adopted \ac{SBOM} formats exist: CycloneDX and the \acf{SPDX}.
Both are built on top of structured and machine-readable data formats, such as JSON and XML, and allow expressing various pieces of information about the product, components, or vulnerabilities (cf. \href{https://cyclonedx.org/capabilities/bov/}{BOV}) and vulnerability exploitability (cf. \href{https://www.ntia.gov/files/ntia/publications/vex_one-page_summary.pdf}{\acs{vex}}).
In their latest versions, both formats are feature equivalent in terms of cybersecurity and vulnerability management, but use different vocabularies. 
For simplicity, we will stick to CycloneDX nomenclature.

\textbf{CycloneDX}\footnote{\url{https://cyclonedx.org/docs/}}, sometimes abbreviated as CDX, is a project of the \ac{OWASP} Foundation started in 2017.
As of writing, the latest version of the standard is \version{1.6}.
An earlier version of CycloneDX is specified in ECMA International standard \ecma{ECMA-424}.

\textbf{System Package Data Exchange (SPDX)}\footnote{\url{https://spdx.dev/specifications}} was proposed by the Linux Foundation in 2011 and is currently available in version \version{3.0.1}.
Originally developed to ensure software license compatibility of dependencies with the host project, it supports a variety of features and use cases today. 
An older version (\version{2.2.1}) is standardized as \iso{ISO/IEC 5962:2021}.

\subsection{Component Identifiers}

At the heart of every \ac{SBOM} is a component list.
To be useful for vulnerability scanning, the \ac{NTIA} stipulates that every component should indicate at least a name, version, and unique identifier~\cite{ntia-sbom-minimal-elements}.
Both common \ac{SBOM} formats, CycloneDX and \ac{SPDX}, are rather permissive and allow adding components (in \ac{SPDX}: \emph{Packages}) without any identifying piece of information.
Although there are multiple means to (uniquely) identify a component, only the following two are frequently used for vulnerability identification.

\textbf{\ac{CPE}} is a \ac{nist} standard format to specify (software, hardware, or operating system) components \cite{nist-cpe}.
A \ac{CPE} is composed of multiple attributes separated by a colon (:).
It usually contains a vendor name, product name, and product version.
In general, attributes can be empty or hold special wildcard characters~\cite{nist-cpe}.
The \ac{CPE} dictionary provides a list of official \ac{CPE} names defining a canonical form of known product and vendor names~\cite{nist-cpe}; the list is regularly updated to include new names and name changes and is publicly available~\cite{nist-cpe-name-dictionary}.
For example, \mbox{\cpe{cpe:2.3:o:linux:linux\_kernel:6.11}} describes Linux Kernel version 6.11. 
Most importantly, the \ac{nvd} uses \ac{CPE} to match known vulnerabilities (identified by their CVE ID) to known products and platforms.

\textbf{\ac{purl}} is a community-driven de facto standard to reference software packages.
Besides the (optional) component name and version, \ac{purl}, requires information on the package manager or ecosystem (e.g., maven, pypi, or npm) a software package resides in;
every \ac{purl} begins with a \enquote{type} specifier (and an optional \enquote{namespace} specifier within the ecosystem).
A minimal \ac{purl} must contain at least a type and a (component) name.
The \ac{purl} \mbox{\code{pkg:deb/debian/linux@6.11}} describes version \version{6.11} of the Linux kernel distributed over the Debian package distribution (deb) as part of the \href{https://www.debian.org}{Debian} Linux Distribution.\\

In contrast to \ac{CPE}, \ac{purl} identifiers can be generated automatically using the ecosystem  and component name.
As a result, especially modern tools prefer \ac{purl} over \ac{CPE}.
However, \ac{purl} can only be used for software components distributed using a package manager, making it impossible to e.g., identify hardware products or source code.
Besides \ac{CPE} and \ac{purl}, several other means to uniquely identify a software component exist, including the \ac{SWID} specified in ISO/IEC~19770-2, the \ac{SWHID}, or the \ac{gitoid}.

\subsection[SBOM-based Vulnerability Scanning]{\acf{svs}}
The \emph{\acf{CVE}} project was born in \citeyear{cve-origins} to allow unique identification of known vulnerabilities by assigning each a standardized ID~\cite{cve-origins}, marking the foundation of modern vulnerability databases.
Today, multiple private and public vulnerability databases and vulnerability identification schemes exist. 
Most importantly, vulnerability databases not only track vulnerabilities, but also link them to affected products using component identifiers such as \ac{CPE} or, more commonly nowadays, \ac{purl}.

In a process we call \emph{\acf{svs}}, a software tool (\svstool{}) processes an \ac{SBOM}, extracts all components (possibly including nested relationships), and then queries vulnerability databases to determine whether known vulnerabilities are associated with any of the components.
In case a vulnerability database returns a positive match, the respective component is known to be affected. 
In organizational setups this may automatically activate an organizational vulnerability management process.

\subsection[Challenges and Limitations of SVS]{Challenges and Limitations of \svs{}}

While \ac{svs} may seem a solved problem on first glance, multiple challenges make automated vulnerability identification difficult in practice.
Most notably, vulnerability identification is a continuous process: components change or get updates, as does public knowledge about vulnerabilities~\cite{vuln-disclosure-delay, vuln-discovery-timeline}. Thus, results are not constant in time.
Furthermore, multiple studies have shown that \ac{SBOM} generators do not create correct and complete \ac{SBOM}s~\cite{impact-of-sbom-generators, creating-sbom-is-hard, sbom_accuracy, challenges-sbom-generation-java, rabbi2024sbom, sbom-svs-study}, thus limiting \ac{svs} capabilities.
For example, creating correct \ac{CPE} component identifiers during \ac{SBOM} generation requires canonicalization of product and vendor names according to the \emph{\ac{CPE} Dictionary} -- for historic reasons, the canonical names are manually selected by the \emph{\acs{CPE} Project}.
Similarly, the data available in vulnerability databases are not always correct and complete~\cite{nvd-quality, nvd-lack-of-data, cvss-inconsistencies} and, in addition, multiple studies show that vulnerability databases publish vulnerabilities with delay \cite{vuln-disclosure-delay}.

Even with perfect \ac{SBOM}s and complete vulnerability databases, vulnerability identification remains challenging.
Matching two \ac{CPE} expressions requires evaluating logical constructs such as wildcards or empty fields~\cite{nist-cpe-matching}, and some \svstool{}s deviate from the official algorithm by limiting wildcard matching.\footnote{\url{https://docs.dependencytrack.org/analysis-types/known-vulnerabilities/}}
Case sensitivity, version ambiguity, and the lack of a standardized versioning schema further complicate comparisons—the \ac{purl} specification treats versions as \enquote{opaque strings}~\cite{purl-spec}.
Since \ac{purl} defines only the identifier format, not the matching logic, and \svstool{}s may also diverge from official \ac{CPE} rules, implementation inconsistencies are inevitable due to diverse parsers, \ac{SBOM} formats, and databases.

Overall, the \ac{svs} process only returns positives results, meaning that no statement about the \enquote{security} of a product or component can be made, if the \svstool{} does not return known vulnerabilities during a lookup.~\footnote{Not all vulnerabilities are known, not all known vulnerabilities are tracked in vulnerability databases (e.g., public disclosure in a blog post), and technical or implementation flaws anywhere along the \ac{svs} process can lead to mismatches and thus false negatives.}

\subsection{Measuring Security}  
Numerous studies tried to analyze factors influence \svstool{} reliability, effectiveness, and by proxy the added security gain of \ac{svs}.
Among others, \citeauthor{impact-of-sbom-generators} show that unreliable quality (i.e., correctness and completeness) of an \ac{SBOM} \cite{creating-sbom-is-hard, sbom_accuracy, sbom-where-we-stand} has major impact on the vulnerability detection capabilities of \svstool{}s~\cite{impact-of-sbom-generators}.
\citeauthor{creating-sbom-is-hard} \cite{creating-sbom-is-hard} presented a benchmark to test and compare \ac{SBOM} generators.
These studies, however, focus on \ac{SBOM} quality and not directly on \svstool{}s.
On the latter, \citeauthor{sbom-svs-study} showed that different \svstool{}s show different results given the same input \ac{SBOM} and that the same \svstool{} report different results when receiving \ac{SBOM}s in a different file formats.
While their results highlight the need for future studies, the authors did not systematically explore the root cause of the observed differences~\cite{sbom-svs-study}.

In general, measuring and quantifying real-world security is a long-standing and difficult problem \cite{Pfleeger2010-MeasuringSecurityIsHard, NIST-MeasuringSecurity, arina-software-security-metrics}.
\citeauthor{anchore-dataset}, creator of the \grype{} \svstool{}, maintains a public dataset of \svstool{} evaluation consisting of container images with corresponding ground-truth labels for expected and unexpected (true-positive and false-positive) CVE~IDs~\cite{anchore-dataset}.
However, this dataset is not based on \ac{SBOM}s and aimed at a specific \svstool{}, making it unsuitable to identify differences between \svstool{}.
In a different setting, \citeauthor{saibersoc}~\cite{saibersoc} propose a test-driven methodology based on injecting handcrafted but realistic signals into the network security monitoring infrastructure of a \acf{SOC} to evaluate its operational performance under real-world conditions.

\smallskip
\textbf{Identified Gap.}
As of today, no method or solution to measure, monitor, and compare the capabilities and maturity of \ac{svs} processes and tools exists.
Without a method to structurally determine failure conditions of \svstool{}s in realistic contexts, the unawareness of unexpected limitations and inconsistencies among \svstool{}s creates a false sense of security whose extend still remains to be determined. 

\section[SVS-Tool Testing Solution (SVS-TEST)]{\svs{}-Tool Testing Solution (\svstts)}
\label{sec:solution}

In this section, we present \svstts{} a method and solution to continuously monitor the vulnerability detection performance of \svstool{}s (and later evaluate its utility in \autoref{sec:evaluation}).
Similar to the methodology presented by \citeauthor{saibersoc}~\cite{saibersoc}, our solution evaluates \svstool{}s using specially crafted evaluation scenarios.
By comparing the output (i.e., the identified vulnerabilities) against the ground truth of each test case, our solution highlights the strengths and limitations of each \svstool{}.
If an \svstool{} does not return the expected results, our solution reveals which combination of input parameters is causing the unexpected or undesirable behavior. 

\svstts{} can be used by organizations building their \svs{} capabilities or to monitor and evaluate the maturity and coverage of existing \svs{} processes.
It allows consumers of \svs{}-as-a-Service solutions (where configuration and behavior can change overnight without the customer's knowledge) to monitor for regressions in the paid service.
Furthermore, \svstool{} developers can use \svstts{} to monitor their own product for regression.

\begin{figure}[tb]
    \centering
    \includegraphics[keepaspectratio,width=1.0\linewidth, clip, trim={0cm 0.4cm 0.1cm 0.19cm}]{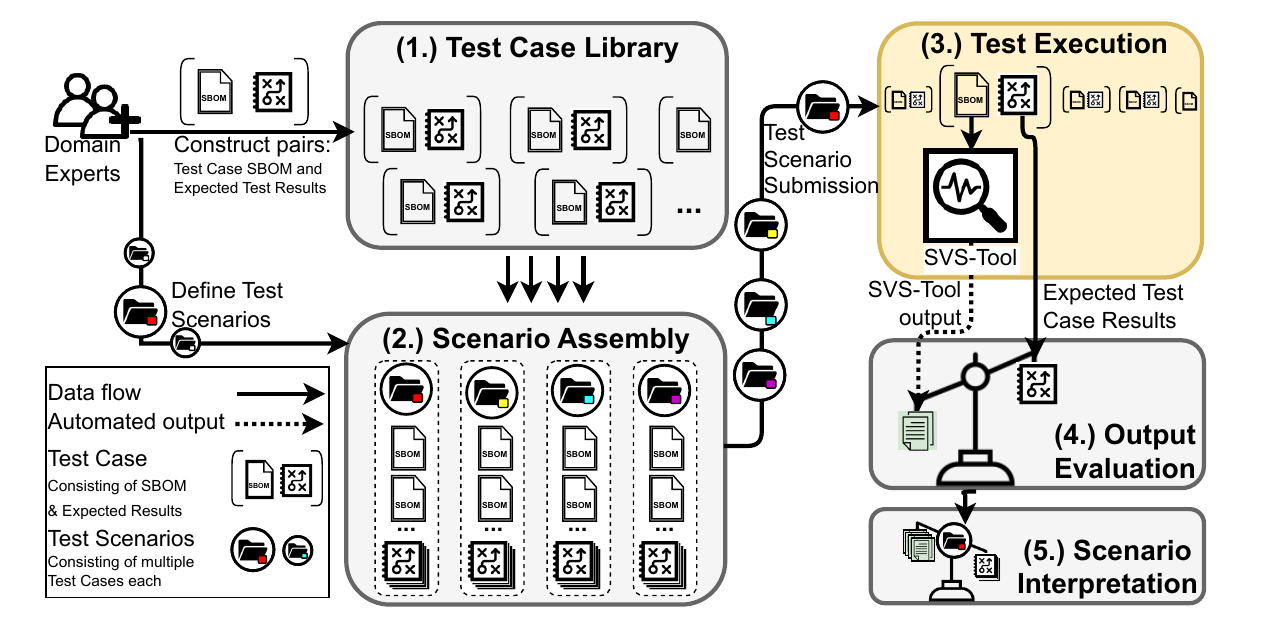}
    \caption{\svs{}-Tool Testing Architecture}
    \label{fig:solution-architecture}
    \Description{
        The Figure visually depicts the Methodology of this work.
        First, domain experts create a Test Case Library in Phase 1. Every entry in the Test Case Library is a tuple consisting of an SBOM and the expected test results.
        In Phase 2, domain experts then group multiple (thematically related) test cases from the Test Case Library into Scenarios.
        During Phase 3, Test Execution, multiple Test Scenarios are run against one ore more SBOM-based Vulnerability Scanning tool (SVS-tool).
        The results of every test case are logged. In Phase 4, Output Evaluation, the obtained results are compared against the expected response.
        Finally, in Phase 5, the individual test case responses are interpreted in scenario context.
        This allows to draw conclusions on the inner workings of the SVS-tool, beyond the results obtained from individual test cases.
    }
\end{figure}

\noindent
Our solution, depicted in \autoref{fig:solution-architecture}, is divided into five `phases': (1.)~test case library, (2.)~scenario assembly, (3.)~test execution, (4.)~output evaluation, and finally (5.)~scenario interpretation.

\paragraph{Test Case Library}  
Each test case consists of two main components, i) the implementation of the test itself, usually in the form of a \ac{SBOM}, and ii) a description of the expected outcome, possible in a machine-readable format (e.g., in form of a regex pattern).
Test cases should include a human-readable description of the creator's intent and a motivation for the expected outcome.
In some cases, expected outcomes are expressed as a list of possible alternatives, such that an expectation is fulfilled if at least one of the alternatives is met.
As an \svstool{} may refuse to process an \ac{SBOM} or crash during execution, error messages or crashes are considered valid expectations when encountering edge case scenarios.

The creators of new test cases should focus on real-world relevance.
Inspiration for test cases can be found in real-world \ac{SBOM}s (e.g., found on the Internet or produced by third-party tools), bugs and limitations identified by a \svstool{} (e.g., GitHub issues), or by trying to enumerate possible interpretation or implementation mistakes \svstool{} developers may have made.  
All test cases are stored in the \emph{test case library}.


\paragraph{Scenario Assembly}  
Expert knowledge is used to design and define \emph{scenarios} using the constructed \emph{test cases}.
In this context, a scenario is defined as a group of individual test cases that are semantically related and focus on evaluating the same general idea, concept, or feature.
Test scenarios should be assembled following a fractional factorial design, so that the exact error conditions can be isolated during \emph{scenario interpretation}.
Although \enquote{incomplete} test scenarios allow drawing meaningful conclusions, test scenarios can and should be expanded with new test cases if necessary.

Each scenario should have a clearly documented expected outcome and human-readable documentation.
In an organizational setting, scenario generation should be guided by organizational \svs{}-capabilities and needs.
For example, an organization committed to SPDX \ac{SBOM}s may not want to run test scenarios implemented for CycloneDX and vice versa.

\paragraph{Test Execution}  
During test execution, an \svstool{} is evaluated using all test cases (of all scenarios).
To avoid confounding or confusion, test cases are submitted individually and the \svstool{} is reset after each test case.
The output generated by the \svstool{}, usually a report containing all identified vulnerabilities is stored for later analysis.
If the \svstool{} produces warning or error messages, these are also stored for later analysis.
Additionally, storing execution timestamps is necessary for regression testing e.g., to see whether an update in the \svs{}-pipeline or \ac{SBOM} generation process leads to a decrease in vulnerability detection capabilities.
For maximum reproducibility, the exact version and configuration of the \svstool{} and a snapshot of all connected vulnerability databases and exact tool configuration and version is necessary, however, not all \svstool{}s provide this information transparently.
The process is repeated for all \svstool{}s under test.


\paragraph[SVS-Output Evaluation]{\svs{}-Output Evaluation}
The stored output from the previous phase (that is, the vulnerability report or error message) is compared against the output expected for the respective test case.
Using regular expressions, our solution can automatically identify whether an expectation is fulfilled or whether the output contains a warning or error message; it distinguishes between three possible outcomes:
\\\textbf{\textit{Pass}} indicates that the expected output was observed, e.g., the output report includes a specific CVE ID.
\\\textbf{\textit{Silent Failure:}} The expected output was not observed and no warning message related to the test case was produced.
This is likely caused by the \svstool{} not supporting the tested feature or functionality, or due to a bug in the \svstool{}.
\\\textbf{\textit{Warning:}} The \svstool{} produced a warning for the given test case. Human insight may be required to determine whether the warning is related to the test case, whether it identifies the error condition, and whether it contains an actionable description on how to fix the problem.
If this is not the case, i.e., because the warning message is unrelated or does not point to the specific problem, the test outcome is set to \emph{silent fail} instead; otherwise, the outcome \emph{warning} remains.

Please note that, in general, passing or failing a test case is not always equivalent to reporting (or failing to report) a vulnerability, but only describes whether the \svstool{} output is according to expectations.
For example, when running a test case with a false positive, one expects \emph{not} to see a specific vulnerability being reported by the \svstool{}.
Similarly, a \emph{silent failure} does not always require action, e.g., if an \svstool{} is very explicit about not supporting a specific feature used in a test scenario.


\paragraph{Scenario Interpretation}  
Finally, by evaluating all test cases of a scenario together, one can conclude on how an \svstool{} works internally or behaves when confronted with a specific scenario.


\section{Case Study}
\label{sec:evaluation}

To evaluate the effectiveness and utility of our solution, we compare \numsvstools{} \svstool{}s using \numtestcases{} test cases grouped into \numtestscenarios{} scenarios.
By doing so, we not only show the need for our solution (e.g., that \svstool{}s behave differently, sometimes unexpectedly) but also demonstrate how our solution can be used in practice to identify and highlight small differences with real-world impact.
All implemented test cases are described in detail in \hyperref[app:test-case-details]{Appendix~\ref*{app:test-case-details}} and our \artifactrepo{}.

Please note that our solution merely highlights the differences between the \svstool{}s;
although it can be used to make direct comparisons between two or more \svstool{}s, our experimental setup is not designed to rate or rank individual tools.
Finding the \enquote{best} \svstool{} for an organization requires the consideration of context factors (e.g., individual requirements, preexisting tooling and processes, \ldots) and is not in the scope of this evaluation.

\subsection{Experimental Variables and Setup}

In our experiment, we define the \svstool{} (including its exact parametrization and configuration, as well as connected vulnerability databases and database version) as \strongemph{independent variable}.
Following our solution architecture, we execute \numsvstools{} experiment runs, one for each \svstool{} with otherwise identical experimental conditions.
During an experiment run, all test cases from all test scenarios are executed (see \autoref{sec:solution}).
To reduce possible confounding factors to a minimum and increase reproducibility, the \svstool{} is reset after each test execution.
As \strongemph{outcome variable}, we record the output of each \svstool{} in response to each test case and map it to either \emph{pass}, \emph{silent fail}, or \emph{warning}, as described in \autoref{sec:solution}.
Despite identical conditions (by the same standardized input test cases), we anticipate seeing notable differences in \svstool{}s output as \strongemph{expected outcome}.
Our \experimentaleval{} is successful, if \svstts{} proves its utility by highlighting differences between \svstool{}s.

\subsection[SVS-Tools]{\svs{}-Tools}
\label{subsec:experiment-svstools}

For our \experimentaleval{}, we selected a total of \numsvstools{} established \svstool{}s and -services.
\svstool{}s were chosen to cover a wide variety of well-known tools and offers, including open source tools, commercial tools, and commercial cloud services.
Four \svstool{}s (\href{https://github.com/devops-kung-fu/bomber}{\bomber{}}, \href{https://github.com/DependencyTrack/dependency-track}{\dtrack{}}, \href{https://github.com/owasp-dep-scan/dep-scan}{\depscan{}}, and \href{https://github.com/anchore/grype}{\grype{}}) are actively maintained open-source projects that can be run locally.
Although the sources for \href{https://github.com/snyk/cli}{\snyk{}} are publicly available, the project no longer accepts community contributions.
The tool is run locally, however, the actual vulnerability lookup happens server-side on the Snyk cloud. 
\dtrack{} and \grype{} are large projects with more than 22 and 16 million downloads respectively (measured on the docker image and GitHub releases).
In contrast, \bomber{} is the youngest project (repository created in July 2022) with, at the time of writing, 21 releases on GitHub and around 0.25 million downloads.
Lastly, we include two (closed source) commercial cloud platforms with subscription plans: \href{https://fossa.com/}{\fossa{}}, and \href{https://vulert.com/}{\vulert{}}. 
For these platforms, we do not have any data on their user base or market penetration, besides what these services self-report on their respective web pages.

All \numsvstools{} \svstool{}s support CycloneDX \ac{SBOM}s in version \version{1.6} and \ac{purl} component identifiers.
Approximately half of the \svstool{}s (\dtrack{}, \grype{}, \fossa{}) also support \ac{CPE} component identifiers.
Most tools, directly or indirectly, access vulnerability data from the \ac{nvd}, \ac{osv}, \acl{oss}, or \ac{ghsa} databases. 
Especially commercial solutions (\snyk{}, \fossa{}, and \vulert{}) may have additional undisclosed vulnerability sources.
\autoref{tab:svstool-version-and-configuration} specifies the exact \svstool{} versions used throughout the \experimentaleval{}.

\subsection{Test Cases and Scenarios}

\begin{table}[tb]
    \centering
    \caption{Overview of Test Scenarios}
    \label{tab:test-case-overview}
    \resizebox{1.0\linewidth}{!}{
    \begin{tabular}{p{0.30\linewidth} p{0.68\linewidth} r} 
    \toprule
        Scenario Name & Scenario Overview & \hspace{-3em} Cases \\ 
    \midrule
        \raggedright 1. Component Identifier Support
          & Baseline to determine basic support for \ac{CPE} and \ac{purl} component identifiers
          & 2
          \\
        \raggedright 2. CPE: No Component Version
          & Test CPE matching compliance for version blank, wildcard (\code{*}), and not applicable (\code{-})
          & 3
          \\
        \raggedright 3. purl: No Component Version
          & Test purl-matching behavior for purl without version information
          & 1
          \\
        \raggedright 4. Component Identifier Priority
          & Test for preferred component identifier if multiple are present (incl. order)
          & 4
          \\
        \raggedright 5. No Component Identifier 
          & Test how \svstool{} behaves if no component identifier is specified in the input
          & 1
          \\
        \raggedright 6. purl: Non-\newline{}canonical prefix
          & Test purl compliance to handle non-canonical (but specification compliant) purl
          & 1
          \\
        \raggedright 7. VEX Ex-\newline{}ploitability Data
          & Test whether CycloneDX VEX data is processed to suppress unrelated vulnerabilities
          & 2
          \\
        \raggedright 8. Invalid Input SBOM
          & Test invalid input SBOM with (i) out-of-order and (ii) unexpected root-level element
          & 2
          \\
    \bottomrule
    \end{tabular}
    }
\end{table}

In total, we implemented and run \numtestcases{} test cases over \numtestscenarios{} scenarios, summarized in \autoref{tab:test-case-overview}.
Test cases are designed to highlight either unexpected behavior of a single \svstool{} on edge-cases or highlight differences between multiple \svstool{}s on the same inputs.
To ensure our test cases have real-world relevance, they were designed based on an examination of relevant standards and \svstool{} bug trackers.
First, we analyzed the specification for CycloneDX, CPE Naming~\cite{nist-cpe}, \ac{CPE} Name Matching~\cite{nist-cpe-matching}, and purl to extract aspects that are either unexpected, believed to be prone to implementation errors, or leave room for (mis)interpretation.
The test cases of \scenario{2,\;3,\;5,\;6, and 8} are the results of this process. 
Secondly, we searched the public GitHub repositories of 5 \svstool{}s for issues documented by the user community.
Ignoring issues that we believe to describe problems specific to a single \svstool{}, we generalized the underlying error condition and created respective test cases.
For example, \scenario{8} is the result of a user describing inconsistencies based on the order of elements in a CycloneDX \ac{BOM}.\footnote{See GitHub issue \href{https://github.com/DependencyTrack/dependency-track/issues/3445}{dependency-track:\#3445}. Note that the CycloneDX specification dictates a fixed order for root-level elements.}
The same GitHub issue also inspired \scenario{4} which also evaluates element order, but applied to a different context within the \ac{SBOM}.
The test cases in \scenario{2,\;4,\;7, and 8} are the result of this process.
For a more detailed rationale (including references to the respective issues), we refer the reader to \appendixref{app:test-case-details}.
An estimate of the impact and how prevalent the test cases are in the real world is provided in \autoref{subsec:discussion-real-world-impact}.

\subsection{Expected Evaluation Results}
Following the results obtained by \citeauthor{sbom-svs-study}~\cite{sbom-svs-study}, we expect to see major differences between \svstool{}s when given the same test case.
Some \svstool{}s exclusively support the purl component identifier format.
If \ac{CPE} \emph{and} \ac{purl} are supported, we expect that \svstool{}s either always prefer one or always use both for vulnerability lookup; we expect their order within the \ac{SBOM} to not have any impact (\scenario{4}).
We expect \svstool{}s to be specification compliant to \ac{CPE} and \ac{purl}, especially with respect to {\scenario{2} and 6} but expect differences where decisions are left to the \svstool{} (\scenario{5}).
We also expect to observe differences in scenarios based on inconsistent or non-standard conform \ac{SBOM}s (\scenario{8}) and scenarios relying on non-essential features (\scenario{7}).
In general, we expect \svstool{}s to produce explicit warnings on error conditions, i.e., never fail silently.
Our expectations are mainly driven by related literature analysis and well-documented inconsistencies in \svstool{} behavior in \svstool{} GitHub issues and in~academic~literature.

\subsection[Experimental Execution \& SVS-Tool Usage]{Experimental Execution \& \svstool{} Usage}
\label{subsec:experimet-implementation}

If not indicated otherwise in \autoref{tab:svstool-version-and-configuration}, we use the default configuration of the \svstool{}s when submitting \ac{SBOM}s for analysis.
We briefly list the most important changes.
\grype{} has a command line option (\code{--add-cpes-if-none}) to enable the reconstruction of \ac{CPE} component identifiers for \ac{SBOM}s that do not contain component identifiers.
Moreover, \acs{vex} processing must be enabled by command line argument (\code{--vex}).
\fossa{} does not perform vulnerability scanning by default; instead, vulnerability checking must be explicitly enabled for each project in the project settings; we activated \enquote{Security Scanning} with the \enquote{Default Security Policy}.
We also manually activated \ac{CPE}-based vulnerability lookup and changed the vulnerability filter to show all found vulnerabilities, as by default, \fossa{} only reports vulnerabilities with severity ranking \enquote{high to critical}.
Lastly, we enabled processing of \enquote{vulnerabilities} from \ac{SBOM} files in the settings.
To use \snyk{}, a (free) account is required.
While Snyk offers paid plans, all our results were obtained using a free account.
Regardless of the configuration, \dtrack{} does not automatically apply \acs{vex} data during \ac{SBOM} import but always requires an additional manual trigger.

\smallskip
Following the method presented in \autoref{sec:solution}, we map all \svstool{} outputs into one of three categories: \emph{pass}, \emph{silent failure}, and \emph{warning}.
Due to space constraints, we describe the exact expectations for each test case in \appendixref{app:test-case-details}.
In most test cases, we expect a specific \acs{CVE}-ID mentioned within the output.
If the expectation value is not observed, we check for unexpected warning and error messages.
We only consider warning messages that are specific to the test case, actionable, and references the affected component inside the \ac{SBOM}.
An accepted warning message could be \enquote{WARN: Component \#1 does not have a testable component identifier}.
In contrast, \enquote{At least one component could not be resolved} is insufficient because it does not identify the affected components and is thus not actionable.
Beware that we do not consider the correctness of the error message, we only evaluate whether it informs the user of an error condition and references the causing component in the \ac{SBOM}.
Log messages are not considered, as they are not presented to the user.
If we neither observe the expected result nor a satisfactory warning message, we set the outcome to \emph{silent failure}.

\boldpseudoparagraph{Example}
Test case \testcase{dmszq6mv} (\scenario{1}) evaluates purl support.
It contains one component, described by a \ac{purl}, that is known to be affected by \cve{CVE-2022-24434}.
We expect all \svstool{}s to report the vulnerability.
A regular expression can search the \svstool{} output for \enquote{\texttt{CVE-2022-24434}}.
Not reporting the vulnerability indicates lack of \ac{purl} support of the \svstool{}.
The test case is \textit{passed} if the regular expression matches; otherwise we use another regular expression to determine the presence of any waning or error messages.
If a warning or error message is present, the test case requires a manual check to ensure that the warning message is relevant and satisfies the requirements.
If it does not, we set the outcome to \textit{silent fail}.
Finally, all remaining outputs are considered \textit{silent failures} and are flagged for manual review.

\begin{table}[tbp]
    \centering
    \caption{Overview of \svstool{} versions and configuration}
    \label{tab:svstool-version-and-configuration}
    \begin{tabular}{p{0.3\linewidth} p{0.2\linewidth} c c}
    \toprule
         \textbf{SVS-Tool} & \textbf{Version} & \textbf{Config.} & \textbf{Cloud?} \\
    \midrule
        \bomber{}  & \version{0.5.1}    &           & \\  
        \dtrack{}  & \version{4.12.0}   &           & \\
        \depscan{} & \version{5.4.6}    &           & \\
        \grype{}   & \version{0.87.0}   & \faPencil & \\
        \snyk{}    & \version{1.1295.2} &           & \faCloudUpload \\
        \fossa{}   & \version{4.27.43}  & \faPencil & \faCloud \\
        \vulert{}  & \version{2.x} *    &           & \faCloud \\
    \bottomrule
    \end{tabular}
    \tablecomment{
        \faPencil{} (some) test cases run with modified configuration. 
        \hspace{0.5em} \faCloud{} \svstool{} is a cloud service. \\
        \faCloudUpload{} \svstool{} is an executable, but scanning is performed server-side in the cloud. \\
        * At the time of writing, \vulert{} does not specify the product version publicly, however a high-ranking contact person told us to refer to it as version \enquote{2.x}.
    }
\end{table}

\section{Results of the Case Study}
\label{sec:results}

In this section, we provide an overview of the results obtained during our \experimentaleval{}. 
For details on every test case, we refer the reader to our \artifactrepo{}.
Please note that the results, summarized in \autoref{tab:results-overview}, mark a snapshot in time.
Since our evaluation, vulnerability databases and \svstool{}s received updates.
In fact, some \svstool{} were updated after we disclosed our test results to the tool vendors (see \nameref{app:appendix}).  
The results in this section therefore no longer reflect the current state-of-the-art, but are primarily designed to show the utility of \svstts{}.

\begin{table*}[tb]
    \newcommand{\pass}[0]{\textcolor{darkgray}{\faCheck}}  
    \newcommand{\fail}[0]{\faRemove}  
    \newcommand{\onlycomp}[0]{\ding{53}}  
    \newcommand{\warn}[0]{\textcolor{darkgray}{\faExclamationTriangle}}  
    \newcommand{\withnote}[0]{$^{f}$}
    \newcommand{\note}[0]{$^{g}$}
    \newcommand{\passwithanote}[0]{$^{a}$} 
    \newcommand{\failbutpass}[1]{\phantom{$^{#1}$}\textcolor{darkgray}{(\faRemove)}$^{#1}$}  
    \newcommand{\notec}[0]{\failbutpass{c}}
    \newcommand{\noted}{\textcolor{darkgray}{(\faCheck)}\textsuperscript{a}} 
    \newcommand{\notes}{\textcolor{darkgray}{(\faCheck)}\textsuperscript{a}} 

    \newcommand{\novex}[0]{\fail$^{b}$}  
    \newcommand{\failbcpe}[0]{\fail$^{c}$}
    \newcommand{\failb}[0]{\fail$^{d}$}
    \newcommand{\failg}[0]{\fail$^{e}$}
    \newcommand{\failgrype}[0]{\fail$^{h}$}
    \newcommand{\rot}[1]{\rotatebox{90}{#1}}
    
    \centering
    \caption{Overview of Results}
    \label{tab:results-overview}
    \begin{tabular}{lll ccccccc}
    \toprule
        Scenario     & Test Case & Description & \rot{Bomber} & \rot{Dep.-Track} & \rot{dep-scan} & \rot{Grype} & \rot{Snyk-CLI} & \rot{FOSSA} & \rot{Vulert}  \\
    \midrule
        \scenario{1} & \testcase{an7esfjj} & CPE support                    & \fail & \pass & \fail & \fail & \warn  & \failbcpe & \fail \\
                     & \testcase{dmszq6mv} & purl support                   & \pass & \pass & \noted & \noted & \pass  & \pass & \pass \\
                     
        \scenario{2} & \testcase{u8h8dnoj} & CPE component version blank    & \fail & \pass & \fail & \warn\note & \warn & \failbcpe & \fail \\
                     & \testcase{fayptrma} & CPE missing component asterisk & \fail & \pass & \fail & \fail & \warn & \failbcpe & \fail \\
                     & \testcase{b5mxq45i} & CPE missing component hyphen   & \fail & \pass & \fail & \fail & \warn & \failbcpe & \fail \\
                     
        \scenario{3} & \testcase{9a7iknu4} & purl without component version & \pass & \fail & \noted & \noted & \fail & \pass & \pass \\
        
        \scenario{4} & \testcase{2lb5zfps} & vulnerable CPE before purl     & \fail & \pass & \fail & \noted & \fail & \fail & \fail \\ 
                     & \testcase{9xhb7rgj} & purl before vulnerable CPE     & \fail  & \pass & \fail & \noted & \fail & \fail & \fail \\
                     & \testcase{pq3cy9or} & CPE before vulnerable purl     & \pass & \pass & \pass & \noted & \pass  & \pass & \pass \\
                     & \testcase{5q46iw4f} & vulnerable purl before CPE     & \pass & \pass & \pass & \noted & \pass  & \pass & \pass \\

        \scenario{5} & \testcase{sqs4tbob} & no component identifier string  & \failb & \fail & \pass & \warn\withnote & \warn & \failg & \fail \\
        
        \scenario{6} & \testcase{hawmnwbz} & non canonical purl string      & \pass & \pass & \pass & \noted & \pass & \pass & \pass \\
        
        \scenario{7} & \testcase{qbqy99do} & CDX VEX with no influence      & \pass & \pass & \pass & \pass & \pass & \pass & \pass \\
                     & \testcase{0vo0efli} & CDX VEX to suppress vuln.      & \fail & \pass & \fail & \warn & \fail & \novex & \fail \\

        \scenario{8} & \testcase{omwcmwv1} & Out of order BOM segments      & \fail & \fail & \fail & \fail & \fail & \fail & \fail \\
                     & \testcase{3fvslnon} & Invalid root-lvl BOM segment   & \fail & \pass & \fail & \fail & \fail & \fail & \fail \\
    \bottomrule
    \end{tabular}
    \tablecomment{
        \pass{}: The expected outcome was observed (e.g., vulnerability was reported suppressed as expected).
        \warn{}: Explicit warning or error message observed.
        \fail{}: Silent failure, i.e., the expected outcome was not observed.
       \texttt{( )}: Test fails due to unrelated incompatibility but the test passes once fixed. 
       \textsuperscript{a}: Test fails when the component version is not explicitly set in an optional CycloneDX field, passes once set.  
       \textsuperscript{b}: Tool recognizes presence of \acs{vex} but does not suppress the respective vulnerability.
       \textsuperscript{c}: Tool extracts components successfully including \ac{CPE}, but the expected vulnerability is not reported.  
       \textsuperscript{d}: The Tool failed to extract components from the \ac{SBOM} and therefore was unable to report the expected vulnerability.  
       \textsuperscript{e}: Components were extracted successfully, but the tool did not report the expected vulnerability.
       \textsuperscript{f}: The Tool supports \ac{CPE} (re)construction from \ac{SBOM} data but does not report the expected vulnerability.
       \textsuperscript{g}: The input \ac{CPE} is incorrectly marked as invalid.

    }
\end{table*}

\paragraph{\scenario{1}: Component Identifier Support}
In general, \fossa{}, \vulert{}, \bomber{}, \snyk{}, and \dtrack{} correctly identified the target components defined in both test cases (one for \ac{CPE}, one for purl), i.e., show the component in their user interface or output.
However, only \dtrack{} correctly reported the expected vulnerability affecting the specified component in both test (\textit{pass}).
\snyk{} does not support \ac{CPE} component identifiers and produced an explicit \textit{warning} message for every untested component \enquote{\code{<Component Name> Info: component must have a purl}}.
Even though \grype{} technically does support \ac{CPE} component identifiers, it does not report the expected vulnerability in the \ac{CPE} test case (\textit{silent fail}).
All other \svstool{}s silently fail the \ac{CPE} test case.
\bomber{}, \snyk{}, \fossa{}, and \vulert{} reported the expected vulnerability in the purl-based test case  (\textit{pass}).
\depscan{} and \grype{} did not include the expected vulnerability in their output.
Manual analysis revealed that both tools do not recognize components that do not have the optional component version field\footnote{\url{https://cyclonedx.org/docs/1.6/json/\#components_items_version}} specified.
Extending our test case \testcase{dmszq6mv} accordingly, makes both tools \textit{pass} the test (highlighted by parentheses in \autoref{tab:results-overview}).

In summary, all \svstool{}s pass the baseline test for purl support but only \dtrack{} passed the baseline test for \ac{CPE} support.
Only \snyk{} (no CPE support) generated an explicit warning for every untested component.
\depscan{} and \grype{} require an (optional) CycloneDX field to be present in the \ac{SBOM}, otherwise they silently fail.
Further investigation revealed that \depscan{} and \grype{} do not directly use the purl for vulnerability search.
Any version information encoded in the \ac{purl} is ignored, instead the tools exclusively considers the optional CycloneDX component \code{version} field of the \ac{SBOM} and silently fails if it is not present. 
Although \fossa{} shows the \ac{CPE} component identifiers extracted from the \ac{SBOM} in its web interface, these are not used for vulnerability analysis.
The user interface does not make it clear that \ac{CPE} identifiers are not included in the vulnerability analysis.
The report technically includes statistics on how many of the identified components have a purl (e.g., \enquote{\texttt{[\ldots]  1/2 components}}) but we neither consider this a warning nor explicit. 


\paragraph{\scenario{2}: Missing Component Version in CPE}
\svstool{} without \ac{CPE} support (see \scenario{1}) also did not report the expected vulnerability in \scenario{2}.
Only \dtrack{} correctly reported the expected vulnerability in all three test cases (blank version/\texttt{ANY}, asterisk/\texttt{ANY}, hyphen/\texttt{NA}) (3x \textit{pass}).
Again, \snyk{} is the only \svstool{} that generates explicit warnings in all three test cases (3x \textit{warning}).
\fossa{} reports the count of components with purl identifiers but we do not consider this an explicit warning, resulting in \textit{silent failure} (x3).
\grype{} successfully extracts the component from the \ac{SBOM} but does not report the expected vulnerability in any of the three test cases.
It \emph{silently fails} (x2) in two occasions, though, When executed with the alternate configuration (\code{--add-cpes-if-none}), it returns an error message for the first test case (version blank), wrongfully claiming that the provided \ac{CPE} is invalid (\texttt{warning}).
As established in \scenario{1}, \bomber{}, \depscan{}, and \vulert{} do not support \ac{CPE} component identifiers and \textit{silently fail}.


\paragraph{\scenario{3}: Missing Component Version in purl}
Only \bomber{}, \fossa{}, and \vulert{} report the expected vulnerability (\textit{pass}); 
\dtrack{} and \snyk{} do not report the expected vulnerability, resulting in a \textit{silent fail}.
Again, \grype{} and \depscan{} fail the original test case due to the missing CycloneDX version field but \textit{pass} it after the adjustment (parentheses in \autoref{tab:results-overview}). 
%
In summary, \dtrack{}, \depscan{}, \grype{} and \snyk{} require optional data and silently fail if not present (\depscan{} and  \grype{}: CycloneDX component version field; \dtrack{} and \snyk{}: version information within the \ac{purl} component identifier.

\paragraph{\scenario{4}: \ac{SBOM} Parser and Component Identifier Order}
Only \dtrack{} correctly reports the vulnerability in all four cases, regardless of whether the vulnerable component is expressed in purl or \ac{CPE} and regardless of the order within the \ac{BOM} file (4x \textit{pass}).
As \bomber{}, \depscan{}, \snyk{}, \fossa{} and \vulert{} do not support vulnerability lookup of \ac{CPE} component identifiers, they only report the expected vulnerability in the two \ac{purl} cases (2x \textit{silent fail}, 2x \textit{pass}).
Again, \grype{} does not report any of the expected vulnerabilities if the optional CycloneDX component version field is not set.
This scenario also showed that \grype{} uses the \ac{purl} from the \ac{SBOM} to report affected components to the user. 
In general, the order of the component identifiers had no impact on any \svstool{}.

\paragraph{\scenario{5}: Missing Component Identifier}
When no component identifier is provided, only \depscan{} reports the expected vulnerability in \scenario{5} (\textit{pass}).
Two \svstool{}s show an explicit warning.
\snyk{} explicitly lists all untested components (error message described in \scenario{1}).
\grype{} also prints a \textit{warning} message \enquote{\code{WARN attempted CPE search on <component name>}} along with the suggestion to re-run the analysis with the \code{--add-cpes-if-none} commandline flag.\footnote{Even with the altered command, \grype{} does not report the expected vulnerability.}
The other \svstool{}s fail silently.

\paragraph{\scenario{6}: Non-canonical purl protocol prefix}
All \svstool{}s correctly process the non-canonical purl string, in compliance with the \ac{purl} specification. 
As established previously, \grype{} only passes if the optional CycloneDX version field is set (otherwise: silent fail).


\paragraph{\scenario{7}: \acsu{vex} Exploitability Analysis}
All \svstool{}s report the expected vulnerability in the baseline case (i.e., all \textit{pass}).
Only \dtrack{} correctly suppressed the vulnerability based on the associated \ac{vex} data in the second test case (\textit{pass}).
Most other \svstool{}s ignore the attached \ac{vex} data and keep reporting the now unexpected (because suppressed) vulnerability (\textit{silent fail}).
\grype{} shows an explicit \emph{warning}: \enquote{\code{unable to find matches against VEX sources}} because it does not support the CycloneDX \ac{vex} format.  
\fossa{} acknowledges the presence of \ac{vex} data in the user interface but does not utilize it.
When using the data export feature, all \ac{vex} data from our test case were overwritten, thus we consider this a \emph{silent fail}.
All remaining tools also \emph{silently fail}.


\paragraph{\scenario{8}: Non-Conforming \ac{BOM} Segments}
When confronted with the (malformed) \ac{BOM} segments out-of-order test case, all tools reported the \enquote{unexpected} vulnerability without any warning (\textit{silent fail}). 
On the second test case (unexpected root-level element), only \dtrack{} rejected the malformed \ac{BOM} with the \textit{warning}: \enquote{The uploaded BOM is invalid, schema validation failed}. 
All other \svstool{}s also \textit{silently failed} the second test case.

\section{Discussion}

In our \experimentaleval{}, we have shown  the utility and practical applicability of \svstts{} using it to find notable differences between \svstool{}s, highlight conditions that lead to unexpected \svstool{} output, and to identify bugs in \svstool{}.

\subsection{Interpretation of Experimental Results}
Most \svstool{}s only support purl and, 
if no \ac{purl} is specified, \svstool{}s show unexpected behavior.
Many \svstool{}s do not even recognize components without a (\ac{purl}), effectively resulting in silently deleting components from the input.
Some \svstool{}s can process \ac{CPE} identifiers, but only use them for vulnerability scanning as a last resort (e.g., \grype{}).
Only \dtrack{} always utilizes all specified identifiers for vulnerability analysis.
In general, our study showed that even if \svstool{}s correctly extracted a component and shows it in the user interface, this does not imply that the component was included in the vulnerability scanning process.
Thus, the information presented on screen may inaccurately reflect the coverage of the security analysis on the \ac{SBOM}.
Interestingly, some tools require a \ac{purl} but do not use it directly for vulnerability scanning.
\grype{} relies on its internal \enquote{grype-db} (a processed aggregation of public vulnerability data) and constructs queries from the component name and version in CycloneDX fields, rather than those in the \ac{purl}.
To the best of our knowledge, \grype{} only uses the \emph{type} field from the \ac{purl} to identify the software ecosystem.
Thus, it can detect vulnerabilities only if the relevant CycloneDX fields and a \ac{purl} are correctly set, but it does not depend on \ac{purl} correctness beyond its type.
This consistent yet poorly documented behavior is also observed in \depscan{}.

The only tool to successfully reconstruct the missing component identifier (in this case, a \ac{CPE}) from the available \ac{BOM} data, perform a vulnerability scan, and then report the expected vulnerability (see \scenario{5}) was \depscan{}.
\grype{} and \snyk{} returned warning messages when components were excluded from the vulnerability analysis, e.g., because of missing or unsupported component identifiers.
All other \svstool{}s silently failed again, creating a false sense of security.
This is especially relevant for projects including hardware components as these cannot be expressed in \ac{purl}.

Only \bomber{}, \fossa{}, and \vulert{} correctly reported the expected vulnerability when the \ac{purl} lacked a version (\scenario{3}), while \dtrack{}, \depscan{}, and \snyk{} silently failed.
Since \ac{purl} (unlike \ac{CPE}~\cite{nist-cpe-matching}) does not define a matching strategy, each \svstool{} implements its own.
\grype{} always relies on the version field and never performs lookups without it.
Overall, the behavior across \svstool{}s is inconsistent; tools may choose to skip lookups to reduce false positives or treat missing versions as wildcards to avoid false negatives, but such policies should be clearly documented.
\snyk{} is the only \svstool{} that reliably showed warning messages when SBOM components lacked the information necessary to perform the vulnerability scan.
Other \svstool{}s mostly fail silently without explicit warning to the user.
This has serious reliability implications, as developers and organizations cannot rely on the completeness of the results obtained from the \svstool{}.
We recommend \svstool{}s to fail explicitly with appropriate warning messages.

\fossa{} lists all components found in an SBOM, with their CPE, purl, and version (from the optional component version field), but vulnerability lookup is only performed on components with a \ac{purl}.
The user interface does not clarify that not all the information displayed on screen is used for vulnerability identification, thus potentially creating a false sense of security.
Although the behavior is consistent, we consider it peculiar and unexpected.

\grype{} depends on the optional component version field and cannot extract components from the SBOM if it is unset.
The authors found this behavior unexpected and suggest issuing a warning.
Developer feedback confirmed that \grype{} uses \ac{CPE} only as a fallback and that our test failures were due to a now-fixed bug.
In one case, \grype{} incorrectly warned about a malformed \ac{CPE}, which we consider a bug.
No \svstool{} had issues with noncanonical \ac{purl} strings in \scenario{6} (\grype{} failed only due to the missing version), and in \scenario{8} all \svstool{}s processed a non-conform CycloneDX BOM without warnings.
Although element order should not affect processing, such BOMs are technically invalid; yet no tool issued warnings.
An open GitHub issue (\href{https://github.com/DependencyTrack/dependency-track/issues/3445}{dependency-track\#3445}) reports a related silent failure in \dtrack{}.

Finally, different \svstool{}s report different vulnerability identifiers for the same vulnerability.
For example, \snyk{} usually refers to vulnerabilities using their own \enquote{SNYK} ID, even if a CVE ID is known and established.
Similarly, other \svstool{}s preferred to report GitHub \ac{ghsa} IDs.

\subsection{Assessment on Real-World Impact}
\label{subsec:discussion-real-world-impact}

During our experimental evaluation, we discovered unexpected behavior and bugs in commonly used flagship \svstool{}s.
Although our test cases are artificial, they are all inspired by real-world problems.
In this section, we perform a preliminary analysis of publicly available \ac{SBOM}s to show that these are not just \enquote{theoretical} problems but that the underlying conditions that trigger unexpected \svstool{} behavior already occur in real-world \ac{SBOM}s.
Thus, in this section, we analyze the \citeyear{sbom-dataset-data} \wildsbomdataset{}~\cite{sbom-dataset-data}, the largest public collection of its kind, featuring \enquote{over 78 thousand unique SBOM files} from \enquote{over 94 million repositories}~\cite{sbom-dataset-data} published by \citeauthor{sbom-dataset-data} as research artifacts to their original work~\cite{sbom-dataset-paper}.
Using the dataset, we identify 772 occasions of \ac{CPE} strings with mixed usage of blank and wildcard values and 7 occurrences of hyphen within a CPE (\scenario{2}).
The dataset also features numerous components with (valid) purl strings without a specified version (\scenario{3}) \cite{sbom-dataset-data}.
We did not observe any non-canonical purl string as described in \scenario{6} within the \wildsbomdataset{}.
Considering that purl strings are likely machine generated, it is unlikely to see non-canonical prefixes in the wild today, even if the specification allows it. That being said, an earlier version of the purl spec did explicitly use the non-canonical prefix.
Although \ac{vex} is actively used in production, e.g., by \href{https://sec.cloudapps.cisco.com/security/center/resources/vex-cvr-faqs}{Cisco}, we did not find a CycloneDX \ac{vex} in the \wildsbomdataset{}.
We did, however, find 7 \ac{SBOM}s using the CycloneDX \emph{vulnerability} section -- the same section in which \ac{vex} data would be located.
We provide further details in \appendixref{app:test-case-details}.

In addition to the publicly available \ac{SBOM}s~\cite{sbom-dataset-data}, we also analyzed 4 CycloneDX (JSON) \ac{BOM}s provided to us by a large European company (ca. 1000 employees, ca. 0.3~billion~Euro annual revenue). 
The company confirmed that the shared \ac{SBOM}s were created by two distinct state-of-the-art commercial service providers that offer vulnerability detection and dependency analysis as a service.
Again, our analysis revealed multiple issues with these \ac{SBOM}s.
In multiple \ac{BOM}s, we found out-of-order root elements, i.e., we frequently discovered the \emph{metadata} section at the end of the document (cf. \scenario{8}).
Furthermore, we observed semantic issues in the metadata, for example, the CycloneDX 1.5 \emph{manufacture} field is supposed to hold information on the manufacturer of the analyzed product described in the \ac{BOM} but we found information on the service provider creating the \ac{BOM} instead.
In few occasions, the metadata specified a natural person as the author of the \ac{SBOM}.
This makes us believe that\ac{SBOM}s were likely machine generated but manually modified afterwards, possibly as part of the provided vulnerability detection service.
In more than 100 occasions, we found components declared without any component identifier (\scenario{5}), although all the data required to construct a \ac{CPE}  (i.e., vendor name, component name, and component version) were present in all instances.
These results show that most (though not all) of our test cases occur in real-world scenarios, albeit with varying likelihood.
The (realistic) assumption that most \ac{SBOM}s are machine generated does not allow to make general statements on the completeness or \enquote{quality} of \ac{SBOM}s in the real-world.
Even \ac{SBOM}s generated by dedicated tools or commercial security service providers are affected by these issues.
We further expect a large hidden figure of low-quality or flawed \ac{SBOM}s not usually accessible to researchers e.g., due to NDAs or enterprise policies.


\subsection{Implications for Practitioners \& Researchers}

Our solution (\svstts{}) and our evaluation results have direct implications for practitioners who actively use \svstool{}s in practice, as well as \svstool{} developers.
Prior research uncovered that \svs{} performance depends on compatibility and quality of the generates \ac{SBOM}s.
In this work, we have shown that these incompatibilities often result in the silent failure of \svstool{}s and that multiple \ac{SBOM}s in the wild meet the identified error conditions.
To ensure the orderly execution of an organizational \ac{svs} process, practitioners should investigate the compatibility between their generated \ac{SBOM}s and \svstool{}s and whether their \svstool{}s are affected by silent failure conditions.
In particular, our results showed that many \svstool{}s require optional CycloneDX fields to operate, do not operate on components without a \ac{purl}, and do not properly communicate the scope of their vulnerability scan to the user.
We believe that our findings have significant impact on organizational \ac{svs} processes, though we also believe that most practitioners are not aware of the identified limitations.

\paragraph{Implications for \ac{svs}-Tool Users}
Our solution \svstts{} provides practitioners with the means to implement custom test cases to evaluate whether their \ac{svs} process is affected by silent failures and thus provides organizations with means to obtain clarity on their \ac{svs} capability and maturity.
This is especially useful, as we have reason to believe that most organizations do not have accurate estimates on the capability, coverage, and maturity of their \svs{} activities (i.e., which components of their products are scanned and which are not).
\svstts{} helps organizations estimate the coverage of their \svs{} process organizations, which is a fundamental requirement for decision making.
In organizational settings with an external service provider (\svs{}-as-a-Service), \svstts{} provides organizations with means to monitor the delivered service.
Our results also have direct impact on organizations working with hardware components, e.g., manufacturers of hardware, or of embedded, IoT, or smart devices.
As these components cannot be expressed in \ac{purl}, affected organizations must choose \svstool{}s that not only support \ac{CPE} component identifiers, but also remain operational in the absence of a \ac{purl}.

\paragraph{Implications for \ac{svs}-Tool Developers}
On multiple occasions, we observed silent failures of \svstool{}s.
In a security context, where the absence of positive results does not allow any conclusions, silent failures are impermissible.
We believe that failures should be explicit (i.e., \emph{not} silent) and that \svstool{}s should provide their users with means to verify correct execution.
For example, \svstool{}s could identify the number of components in the input \ac{SBOM}s, track which components were used during vulnerability discovery, and report any mismatch.
Ideally, warning messages should contain a clear identification of any component in the \ac{SBOM} that was was not included in the vulnerability identification phase, as well as information why. Example: \enquote{Component \textless{}\mbox{bomRef}\textgreater{} excluded: no purl}.
While implementing explicit notifications does not improve the capability of an \svstool{}, it provides all \svstool{} users with means to monitor and improve their \ac{svs} process.
Furthermore, we believe that \svstool{}s should be more explicit about their assumption and failure conditions.
We did not find explicit documentation on which component data is required for the \svstool{} for the vulnerability identification process for any of the tested \svstool{}s.
\svstool{} developers can easily address this shortcoming by updating their documentation.
Finally, we have shown that \svstts{} is successful in identifying real-world bugs and regressions.
\svstool{} developers can utilize our method to create their own battery of tests to check for regression in new releases.
It is difficult to estimate how many organizations (or people) were affected by our findings, thus the exact security impact remains unknown.
Considering that \svstool{}s are used by multiple million users or organizations for multiple projects with thousands of components each, even a small bug may have significant real-world impact.
The possible impact scales drastically when consumer products are involved (e.g., IoT and smart devices).
In these cases, an unrecognized security issue may (unnecessarily) put millions of customers across the entire globe at risk.
Our analysis showed that \ac{SBOM}s causing these types of silent failures exist in the real-world.
Considering that we also found multiple indicators of regression in public \svstool{} GitHub issue trackers, we believe that \svstool{} developers and the \ac{svs} community as a whole would collectively benefit from such efforts.

\paragraph{Implications for Researchers}
Our results show that, in practice, \svstool{} behavior may deviate from expectations when given seemingly normal \ac{SBOM}s.
While performed an initial impact estimation based on public \ac{SBOM}s in the wild, further research is necessary to quantify the real-world impact in practice.
When studying an organization's security management process, researchers can use \svstts{} to measure and study the differences between the expected component coverage and the coverage actually achieved by the \ac{svs} process.
Because \svstts{} is designed around tests against a ground-truth, it enables researchers to develop experiments to measure the impact of false-negatives on organizational security, something that is otherwise difficult to study in a real-world context.
Furthermore, \svstts{} allows researchers to study the actionability and user-friendlessness of \svstool{}s, their vulnerability reports, and error messages -- especially because prior research identified the lack in usability as a limiting factor to \ac{SBOM} adoption~\cite{sbom-on-github-projects, sbom_developer, sbom-svs-study, linux-foundation-sbom-readiness-survey}.



\subsection{Ethical Considerations \& Disclosure Process}
\label{subsec:ethical-considerations}
\svstts{} highlights limitations in vulnerability detection in \svstool{}s.
As such, our results do not directly threaten the cybersecurity of any organization or \svstool{} and do not require vulnerability disclosure.
We contacted all involved \svstool{} developers prior to the submission and independently disclosed all our findings related to their respective \svstool{}.
If the email address of key personnel was not known from a prior exchange, we searched for a dedicated contact email address or identified the project lead or maintainer of the GitHub repository.
In multiple occasions, we opened support tickets on GitHub, most of which were already addressed by the \svstool{} developer by the time of writing.
An overview of disclosed bugs and feedback is available in the \nameref{app:appendix}.  

\section{Conclusion}
We presented \svstts{}, a solution to systematically monitor and evaluate \svstool{} behavior and to practically determine capabilities, maturity, and failure conditions of \svstool{}s in a real-world environment.
By sharing this tool with researchers and practitioners, we address an open research gap~\cite{sbom-svs-study, sbom_developer, sbom-on-github-projects}.
We showed the utility of \svstts{} in a \experimentaleval{} by identifying multiple failure conditions and bugs in real-world \svstool{}s. 
We showed that \svstts{} is capable of identifying inconsistencies, unexpected behavior, and bugs and contributed to improving \svs{} maturity as a whole.
Due to its customizability, \svstts{} can be used by researchers and practitioners alike, e.g., to monitor \ac{svs} capabilities and maturity in organizations or in \svstool{} development.
Further work may explore means to automate scenario and test case creation and evaluation.
While we have shown that even established \svstool{}s produce unexpected results in response to valid \ac{SBOM}s, more work is necessary to quantify the impact and likelihood.




\begin{acks}
    The project is funded in equal parts by the University of Padua and VIMAR S.p.A. under project \enquote{AISec4IoT}.
We thank the commercial \svstool{} vendors who provided us with a free license.
For the purpose of open access, a CC BY 4.0 public copyright license is applied to any Author Accepted Manuscript arising from this submission.
Author contribution statements \enquote{CRediT} available (see \nameref{app:appendix}).
\end{acks}

\printbibliography

@inproceedings{sbom-svs-study,
  title = {Impacts of Software Bill of Materials (SBOM) Generation on Vulnerability Detection},
  doi = {10.1145/3689944.3696164},
  booktitle = {Softw. Supply Chain Offens. Res. and Ecosyst. Def.},
  _publisher = {ACM},
  author = {O’Donoghue,  Eric and Boles,  Brittany and Izurieta,  Clemente and Reinhold,  Ann Marie},
  year = {2023},
  date = {2023-11},
}

@article{nvd-quality,
  author={Anwar, Afsah and Abusnaina, Ahmed and Chen, Songqing and Li, Frank and Mohaisen, David},
  journal={Trans. on Dependable and Secur. Comput.}, 
  title={Cleaning the {NVD}: Comprehensive Quality Assessment, Improvements, and Analyses},
  _publisher = {IEEE},
  year={2022},
  volume={19},
  _number={6},
  _keywords={Security;Databases;Standards;Reliability;Market research;Tools;Systematics;Vulnerability analysis;CVSS;NVD},
  doi={10.1109/TDSC.2021.3125270}
}

@techreport{linux-foundation-sbom-readiness-survey,
    author = {Stephen Hendrick},
    title = {Software Bill of Materials (SBOM) and Cybersecurity Readiness},
    institution = {The Linux Foundation},
    year = {2022},
    month = {01},
    date = {2022-01},
    _note = {foreword by Jim Zemlin},
    url = {https://www.linuxfoundation.org/research/the-state-of-software-bill-of-materials-sbom-and-cybersecurity-readiness},
    urldate = {2024-12-12},
}

@article{sbom-on-github-projects,
  author = {Bi,  Tingting and Xia,  Boming and Xing,  Zhenchang and Lu,  Qinghua and Zhu,  Liming},
  title = {On the Way to SBOMs: Investigating Design Issues and Solutions in Practice},
  year = {2024},
  doi = {10.1145/3654442},
  journal = {Trans. on Softw. Eng. and Methodol.},
  publisher = {ACM},
  _issn = {1557-7392},
  volume = {33},
  _number = {6},
  _month = {6},
  date = {2024-06},
}

@inproceedings{sbom_accuracy,
    author = {Halbritter, Andreas and Merli, Dominik},
    title = {Accuracy Evaluation of SBOM Tools for Web Applications and System-Level Software},
    year = {2024},
    _publisher = {ACM},
    doi = {10.1145/3664476.3670926},
    booktitle = {Int. Conf. on Availab., Reliab. and Secur.},
    volume = {13},
    _series = {ARES '24}
}

@INPROCEEDINGS{sbom_developer,
  author={Otoda, Wataru and Kanda, Tetsuya and Manabe, Yuki and Inoue, Katsuro and Higo, Yoshiki},
  title={SBOM Challenges for Developers: From Analysis of Stack Overflow Questions}, 
  booktitle={Softw. Eng. Res., Manag. and Applica.},   
  _booktitle={Int. Conf. on Softw. Eng. Res., Manag. and Applica.}, 
  year={2024},
  _keywords={Bills of materials;Software;Libraries;Security;Software engineering;Software development management;SBOM;SPDX;CycloneDX;Software Supply Chain;Stack Overflow},
  doi={10.1109/SERA61261.2024.10685624}
}

@report{nist-cpe,
    author = {Cheikes, Brant A. and Waltermire, David and Scarfone, Karen},
    title = {Common Platform Enumeration: Naming Specification Version 2.3},
    year = {2011},
    date = {2011-08-19},
    doi = {10.6028/NIST.IR.7695},
    _institution = {National Institute of Standards and Technology},
    _number = {Interagency Report 7695},
    note = {NIST Interagency Report 7695},
}

@report{nist-cpe-matching,
    author = {Parmelee, Mary and Booth, Harold and Waltermire, David and Scarfone, Karen},
    title = {Common Platform Enumeration: Name Matching Specification Version 2.3},
    year = {2011},
    date = {2011-08-19},
    doi = {10.6028/NIST.IR.7696},
    _institution = {National Institute of Standards and Technology},
    _number = {Interagency Report 7696},
    note = {NIST Interagency Report 7696},
}

@report{onekey,
    author = {{Onekey GmbH}},
    title = {IoT \& OT Cybersecurity Report},
    year = {2025},
    url = {https://www.onekey.com/resource/iot-ot-cybersecurity-report-2025},
    urldate = {2025-12-12}
}

@dataset{nist-cpe-name-dictionary,
    author = {{NIST CPE Project}},
    title = {Official {CPE} Dictionary},
    year = {},
    url = {https://nvd.nist.gov/products/cpe},
    urldate = {2025-10-10},
}

@misc{purl-spec,
    key = {purl specification},
    title = {Purl Specification},
    year = {2017},
    date = {2017-11-11},
    url = {https://github.com/package-url/purl-spec},
    urldate = {2025-10-09},
}

@misc{owasp_risk,
  author       = {OWASP},
  title        = {Top 10 Open Source Software Risks},
  year         = {2024},
  url = {https://owasp.org/www-project-open-source-software-top-10/},
  urldate = {2025-01-27}
}

@misc{owasp-top-ten-2013,
  author       = {OWASP},
  title        = {Top 10 Web Application Risks},
  year         = {2013},
  url = {https://github.com/OWASP/Top10/tree/master/2013},
  _url = {https://github.com/OWASP/Top10/blob/master/2013/OWASP%20Top%2010%20-%202013.pdf},
  urldate = {2025-02-10}
}

@article{impact-of-sbom-generators,
  author = {Benedetti,  Giacomo and Cofano,  Serena and Brighente,  Alessandro and Conti, Mauro},
  title = {The Impact of SBOM Generators on Vulnerability Assessment in Python: A Comparison and a Novel Approach},
booktitle = {23rd International Conference on Applied Cryptography and Network Security (ACNS)},
  year = {2025},
  _doi = {10.48550/ARXIV.2409.06390},
  eprinttype = {arXiv},
  eprint = {2409.06390},
  archivePrefix={arXiv},
  primaryClass={cs.SE},
}

@article{vuln-discovery-timeline,
  author = {Yi Wen Heng and Zeyang Ma and Haoxiang Zhang and Zhenhao Li and  Tse-Hsun (Peter) Chen},
  title = {Discovery of Timeline and Crowd Reaction of
Software Vulnerability Disclosures},
  year = {2024},
  eprinttype = {arXiv},
  eprint = {2411.07480},
  archivePrefix={arXiv},
  primaryClass={cs.SE},
}

@inproceedings{vuln-disclosure-delay,
 author = {Luis Gustavo Araujo Rodriguez and Julia Selvatici Trazzi and Victor Fossaluza and Rodrigo Campiolo and Daniel Macêdo Batista},
 author = {Luis G. A. Rodriguez and Julia Selvatici Trazzi and Victor Fossaluza and Rodrigo Campiolo and Daniel Macêdo Batista},
 title = {Analysis of Vulnerability Disclosure Delays from the National Vulnerability Database},
 booktitle = {Segur. Cibern. em Disposit. Conectados},
 _location = {São José dos Campos},
 year = {2018},
 _publisher = {SBC},
 _address = {Porto Alegre, RS, Brasil},
 url = {https://sol.sbc.org.br/index.php/wscdc/article/view/2394}
}

@inproceedings{cvss-inconsistencies,
  author = {Zhang,  Siqi and Cai,  Minjie and Zhang,  Mengyuan and Zhao,  Lianying and de Carnavalet,  Xavier de Carné},
  title = {The Flaw Within: Identifying CVSS Score Discrepancies in the NVD},
  doi = {10.1109/cloudcom59040.2023.00039},
  _publisher = {IEEE},
  booktitle = {Int. Conf. on Cloud Comput. Technol. and Sci.},
  year = {2023},
  date = {2023-12},
}

@inproceedings {nvd-lack-of-data,
    author = {Dongliang Mu and Alejandro Cuevas and Limin Yang and Hang Hu and Xinyu Xing and Bing Mao and Gang Wang},
    title = {Understanding the Reproducibility of Crowd-reported Security Vulnerabilities},
    booktitle = {27th USENIX Security Symposium},
    year = {2018},
    _isbn = {978-1-939133-04-5},
    url = {https://www.usenix.org/conference/usenixsecurity18/presentation/mu},
    _publisher = {USENIX Association},
    date = {2018-08},
}

@report{ntia-sbom-minimal-elements,
  author = {{National Telecommunications and Information Administration (NTIA)}},
  title = {The Minimum Elements For a Software Bill of Materials ({SBOM})},
  year = {2021},
  date = {2021-07-12},
  url = {https://www.ntia.doc.gov/files/ntia/publications/sbom_minimum_elements_report.pdf},
  urldate = {2025-02-11}
}

@dataset{anchore-dataset,
    author = {{Anchore Inc.}},
    title = {vulnerability-match-labels},
    year = {2022},
    date = {2022-09-21},
    url = {https://github.com/anchore/vulnerability-match-labels},
    urldate = {2025-02-15},
    _organization = {Anchore},
}

@inproceedings{sbom,
    author = {Dalia, Gregorio and Visaggio, Corrado Aaron and Di Sorbo, Andrea and Canfora, Gerardo},
    title = {SBOM Ouverture: What We Need and What We Have},
    year = {2024},
    _publisher = {ACM},
    doi = {10.1145/3664476.3669975},
    booktitle = {ARES},
    volume = {19},
    keywords = {Open Source Software, Software Bill of Materials, Software Supply Chain},
}

@inproceedings{challenges-sbom-generation-java,
  author = {Balliu,  Musard and Baudry,  Benoit and Bobadilla,  Sofia and Ekstedt,  Mathias and Monperrus,  Martin and Ron,  Javier and Sharma,  Aman and Skoglund,  Gabriel and Soto-Valero,  César and Wittlinger,  Martin},
  title = {Challenges of Producing Software Bill of Materials for Java},
  year = {2023},
  doi = {10.1109/msec.2023.3302956},
  journal = {IEEE S\&P},
  issn = {1558-4046},
  volume = {21},
  _number = {6},
  date = {2023-11},
}

@techreport{cve-origins,
    author = {Mann, David E. and Christey, Steven M.},
    title = {Towards a Common Enumeration of Vulnerabilities},
    institution = {MITRE},
    year = {1999},
    date = {1999-01-08},
    url = {https://www.cve.org/Resources/General/Towards-a-Common-Enumeration-of-Vulnerabilities.pdf},
    urldate = {2025-10-08},
}

@article{saibersoc,
  author = {Rosso, Martin and Campobasso, Michele and Gankhuyag, Ganduulga and Allodi, Luca},
  title = {SAIBERSOC: A Methodology and Tool for Experimenting with Security Operation Centers},
  year = {2022},
  doi = {10.1145/3491266},
  journal = {ACM DTRAP},
  _journal = {Digital Threats: Research and Practice},
  _volume = {3},
  _issn = {2576-5337},
  date = {2022-02},
}

@inproceedings{sbom-where-we-stand,
  title = {An Empirical Study on Software Bill of Materials: Where We Stand and the Road Ahead},
  doi = {10.1109/icse48619.2023.00219},
  booktitle = {Int. Conf. on Softw. Eng.},
  volume = {45},
  _publisher = {IEEE},
  author = {Xia,  Boming and Bi,  Tingting and Xing,  Zhenchang and Lu,  Qinghua and Zhu,  Liming},
  year = {2023},
  _date = {2023-05} 
}

@article{Pfleeger2010-MeasuringSecurityIsHard,
  title = {Why Measuring Security Is Hard},
  author = {Pfleeger, Shari Lawrence and Cunningham, Robert K},
  year = {2010},
  doi = {10.1109/msp.2010.60},
  issn = {1540-7993},
  volume = {8},
  _number = {4},
  journal = {IEEE S\&P},
  month = {07},
  date = {2010-07},
}

@techreport{NIST-MeasuringSecurity,
   author = {Wayne Jansen},
   title = {Directions in Security Metrics Research},
   year = {2009},
   % institution = {National Institute of Standards and Technology},
   _number = {Interagency Report 7564},
   note = {NIST Interagency Report 7564},
   url = {https://nvlpubs.nist.gov/nistpubs/legacy/ir/nistir7564.pdf},
   urldate = {2024-12-18},
}

@inproceedings{arina-software-security-metrics,
    author = {Kudriavtseva, Arina and Gadyatskaya, Olga},
    title = {You cannot improve what you do not measure: A triangulation study of software security metrics},
    year = {2024},
    doi = {10.1145/3605098.3635892},
    booktitle = {Symposium on Applied Computing},
    pages = {1223–1232},
    _series = {SAC '24}
}

@inproceedings{creating-sbom-is-hard,
  author = {Yu,  Sheng and Song,  Wei and Hu,  Xunchao and Yin,  Heng},
  title = {On the Correctness of Metadata-Based SBOM Generation: A Differential Analysis Approach},
  year = {2024},
  date = {2024-06},
  doi = {10.1109/dsn58291.2024.00018},
  booktitle = {Int. Conf. on Dependable Syst. and Netw.},
  _publisher = {IEEE},
  _volume = {54},
}

@inproceedings{rabbi2024sbom,
    title = {SBOM Generation Tools Under Microscope: A Focus on The npm Ecosystem},
    author = {Rabbi, Md Fazle and Champa, Arifa Islam and Nachuma, Costain and Zibran, Minhaz Fahim},
    booktitle = {Symposium on Applied Computing},
    year = {2024},
    doi = {10.1145/3605098.3635927},
    _abstract = {Generating accurate Software Bill of Materials (SBOM) is challenging due to the complex dependencies in the diverse components used in software and also the way software is built into executables. A handful of tools claim the capability of automatic SBOM generation from software distributions while little is known about their applicability, strengths, and limitations. Our study makes quantitative and qualitative comparisons of the four such tools (i.e., ORT, cnn, syft, cdxgen) that claim to be capable of generating SBOM from JavaScript projects. For the comparison, we operate these four tools on 50 open-source JavaScript npm projects. We find significant performance variations when evaluating their ability to extract component details, especially in detecting dependencies. The findings of this study are useful in the design and development of SBOM generator tools, in end-users' selections of such tools, and thus in the overall improvement of the security and transparency in software supply chain.},
}

@misc{sbom-dataset-paper,
  Author = {Luıs Soeiro and Thomas Robert and Stefano Zacchiroli},
  Title = {Wild SBOMs: a Large-scale Dataset of Software Bills of Materials from Public Code},
  Year = {2025},
  Eprint = {2503.15021},
  eprinttype = {arXiv},
  eprint={2404.17955},
  archivePrefix={arXiv},
  primaryClass={cs.SE},
  note = {In Press},
}

@dataset{sbom-dataset-data,
  doi = {10.5281/ZENODO.14250102},
  author = {Soeiro,  Luis and Robert,  Thomas and Zacchiroli,  Stefano},
  title = {Replication Package for Wild SBOMs: a Large-scale Dataset of Software Bills of Materials from Public Code},
  publisher = {Zenodo},
  year = {2024},
  copyright = {CC BY 4.0}
}

\appendix
\section{Appendix}
\label{app:appendix} 
Appendix available from the \href{https://doi.org/10.5281/zenodo.17921254}{\artifactrepo{}}.



\subsection{Test Case Details}
\label{app:test-case-details}

In this section, we provide details on the most relevant characteristics of each test case, including their motivation and an initial assessment on how common or prevalent these edge cases are in real world \ac{BOM}s.
Please find test case implementations in our \artifactrepo{}.

In most cases, we expect the \svstool{} to report a specific vulnerability in response to the test case.
In such cases we specify the \ac{CVE} referencing the expected vulnerability, however during the test any equivalent identifier referencing the same vulnerability is accepted.
When an \svstool{} does not show the expected behavior, we consider this a \textit{silent failure} unless the \svstool{} shows an explicit info or warning message to the end user stating that test results may be incomplete due to an error condition.
To consider a warning message satisfactory, we require it to be actionable and clearly reference the affected component inside the \ac{BOM}.
An example for an accepted warning message is \enquote{WARN: Components  \#1, \#2, \#9 do not have a testable component identifier}.
In contrast, a warning message such as \enquote{At least one component could not be resolved} is not actionable as it does not help identify which component is affected and what caused the problem.
We also consider faulty error messages, as long as they correctly reference the affected data field in the affected component.

\medskip

\testcasetitle{an7esfjj}{Scenario 1}

\subparagraph{Summary} 
\ac{CPE} support in the SBOM.

\subparagraph{Description}
This test case features a valid \ac{CPE} with an attribute \cpe{cpe:2.3:a:vendor:component:version:*\\:\ldots{}}. The described component \componentp{npm}{dicer} in these test cases is affected by \cve{CVE-2022-24434}.

\subparagraph{Rationale}
There are several types of identifiers, but the two most widely used are \ac{CPE} and \ac{purl}. However, not all \svstool{}s support every type of identifier; in fact, some \svstool{}s do not support \ac{CPE}. 

\subparagraph{Motivation}
Because the \ac{purl} specification does not support hardware components, especially \ac{BOM}s for embedded and IoT platforms opt for \ac{CPE} component identifiers.
For example, the Open-source \href{https://github.com/espressif/esp-idf-sbom}{Espressif IDF SBOM} generator tool\footnote{Espressif is a large manufacturer for microcontrollers used in IoT} solely focuses on \ac{CPE} component identifiers. 

\subparagraph{Expectation}
We expect that all \svstool{} pass the test case by reporting the expected vulnerability \cve{CVE-2022-24434}.
Alternatively, we expect \svstool{}s that do not support \ac{CPE} component identifiers to explicitly warn the user that the component in  our test case was not scanned for known vulnerabilities because no supported component identifier was found.


\medskip

\testcasetitle{dmszq6mv}{Scenario 1}

\subparagraph{Summary} 
\ac{purl} support in the SBOM.

\subparagraph{Description}
This test case features a valid \ac{purl} with an attribute \purl{pkg:type/name@version}. The described component \componentp{npm}{dicer} in these test case is affected by \cve{CVE-2022-24434}.

\subparagraph{Rationale}
The two most commonly used identifiers are \ac{CPE} and \ac{purl}. However, not all \svstool{}s support both. Some tools work only with \ac{purl}, while others can handle both \ac{purl} and \ac{CPE}.

\subparagraph{Motivation} 
We use \ac{SBOM}s that include \ac{purl} identifiers; examples of such \ac{SBOM}s can be found in the CycloneDX \href{https://github.com/CycloneDX/bom-examples/tree/master/SBOM}{BOM-Examples} GitHub repository. A well-known \ac{SBOM}-generating tool \href{https://github.com/anchore/syft}{Syft} produces \ac{SBOM}s that include both \ac{purl} and \ac{CPE} identifiers. Examples of \ac{SBOM}s generated by Syft for Docker Hub container images are available in their \href{https://github.com/anchore/sbom-examples}{GitHub repository}.

\subparagraph{Expectation}
These are simple test cases; therefore, we expect that a \svstool{} utilizing \ac{purl} identifiers for vulnerability matching should correctly identify and report the expected vulnerability, in our case it is \cve{CVE-2022-24434}.


\medskip

\testcasetitle{u8h8dnoj}{Scenario 2}

\subparagraph{Summary}
\ac{CPE} component version left blank.

\subparagraph{Description}
This test case features a valid \ac{CPE} with a blank version attribute \cpe{cpe: \ldots{} :: \ldots{}}.
The described component \componentp{npm}{dicer} in these test cases is affected by \cve{CVE-2022-24434}.

\subparagraph{Rationale}  
\ac{SBOM} generators may not always be able to accurately determine the exact version of a component. During the generation process, internal \texttt{null} values may be serialized as empty strings, or alternatively represented using the special \enquote{\code{*}} symbol. Both approaches are valid and commonly observed in practice. In fact, we identified a total of 772 \ac{CPE} entries with blank and asterisk version fields within the \wildsbomdataset{}~\cite{sbom-dataset-data}.

\subparagraph{Motivation}  
The \ac{CPE} specification clearly states that blank values are to be interpreted as logical \texttt{ANY}~\cite{nist-cpe} and must always match~\cite{nist-cpe-matching}. To produce better results, \svstool{}s do not implement \ac{CPE}-matching exactly as specified in the standard~\cite{nist-cpe-matching}. In fact, several tools implement a custom variation of the matching algorithm; for example, fuzzy matching is used in \dtrack{}, as discussed in \href{https://github.com/DependencyTrack/dependency-track/discussions/2983}{Discussion \#2983}, while \grype{} provides configurable matching behavior as mentioned on their \href{https://github.com/anchore/grype#configuration}{GitHub repository}. Furthermore, there are multiple instances of flawed implementations across various \svstool{}s, such as the case in \dtrack{} \href{https://github.com/DependencyTrack/dependency-track/issues/2894}{GitHub Issue \#2894}.

\subparagraph{Expectation}
We expect all \svstool{}s to correctly process \ac{CPE}s with an empty version string.
We consider a test passing, if the \svstool{} reports the expected vulnerability known as \cve{CVE-2022-24434}.
If the \svstool{} does not report the expected vulnerability but also does not explicitly inform the user about possible error states, we consider this a silent failure. 
Examples for accepted warning messages are \enquote{Error processing CPE for component X}, or \enquote{Component Y: Invalid CPE} (even though the \ac{CPE} is standard conform).

\medskip
\testcasetitle{b5mxq45i}{Scenario 2}

\subparagraph{Summary}
\ac{CPE} with component version as a hyphen.

\subparagraph{Description}
This test case features a valid \ac{CPE} with a hyphen in the version attribute \cpe{cpe: \ldots{} :-: \ldots{}}.
The described component \componentp{npm}{dicer} is affected by \cve{CVE-2022-24434}.

\subparagraph{Rationale} \ac{SBOM} generators may produce \ac{CPE} entries containing a hyphen (\enquote{-}) to indicate special value \enquote{Not Applicable (NA)}.
This value shall only be used of the respective \ac{CPE} attribute does not have any meaningful value e.g., the language attribute does not have any meaningful value for a hardware product.
An example of this issue was highlighted in the \href{https://github.com/intel/cve-bin-tool/issues/481}{GitHub Issue \#481}, where a user reported that their \ac{SBOM} included a \ac{CPE} entry with a hyphen. However, using \enquote{NA} as a component version in an identifier is not meaningful; it only makes sense when specifying affected versions in a vulnerability database.
\subparagraph{Motivation} The \ac{CPE} specification clearly states that hyphen values are to be interpreted as logical \texttt{NA}~\cite{nist-cpe}. We identified a total of 7 \ac{CPE} entries with hyphens in the \cpe{CPE} attributes, within the \wildsbomdataset{}~\cite{sbom-dataset-data}.%

\subparagraph{Expectation} We expect all \svstool{}s to correctly process \ac{CPE}s with a hyphen version string.
We consider a test passing, if the \svstool{} reports the expected vulnerability, \cve{CVE-2022-24434}.
If the \svstool{} does not report the expected vulnerability but also does not explicitly inform the user about possible error states, we consider this a silent failure. 
Examples for accepted warning messages are \enquote{Error processing CPE for component X}, or \enquote{Component Y: Invalid CPE} (even though the \ac{CPE} is standard conform).

\medskip

\testcasetitle{fayptrma}{Scenario 2}

\subparagraph{Summary}
\ac{CPE} component version with a asterisk.

\subparagraph{Description}
This test case features a valid \ac{CPE} with an asterisk in the version attribute \cpe{cpe:\ldots{}:*:\ldots{}}.  
The component described as \componentp{npm}{dicer} in these test cases is affected by \cve{CVE-2022-24434}.

\subparagraph{Rationale} 
Identical to the test case \testcase{u8h8dnoj}.

\subparagraph{Motivation}  
The \ac{CPE} specification clearly states that asterisk values are to be interpreted as logical \texttt{ANY}~\cite{nist-cpe} and must always match~\cite{nist-cpe-matching}. Syft can produce \ac{CPE}s with asterisks, as \texttt{*} means \texttt{ANY}. This can lead to false positives; one such issue was highlighted in Syft \href{https://github.com/anchore/syft/issues/396}{GitHub Issue \#396}. 

\subparagraph{Expectation}
Identical to the test case \testcase{u8h8dnoj}.

\medskip

\testcasetitle{9a7iknu4}{Scenario 3}

\subparagraph{Summary}
The version specifier is absent from the \ac{purl}.

\subparagraph{Description}
This test case features a \ac{purl} without the version in the \ac{purl} string \purl{pkg:ecosystem/package}.  
The component described as \componentp{npm}{dicer} in this test case is affected by \cve{CVE-2022-24434}.

\subparagraph{Rationale} 
The version field in a \ac{purl} is considered optional, as specified in the \href{https://github.com/package-url/purl-spec}{purl specification}.
We identified 22\;320 \ac{purl} entries without version information in the \wildsbomdataset{}~\cite{sbom-dataset-data}, one of the largest publicly accessible databases of \ac{SBOM}s. 

\subparagraph{Motivation}  
The \ac{purl} specification states that the \texttt{version} field is optional, while the \texttt{scheme}, \texttt{type}, and \texttt{name} are required fields. We observed a \dtrack{} \href{https://github.com/DependencyTrack/dependency-track/issues/1115}{GitHub Issue \#1115} where such scenarios exist for \svstool{}s, where it can be put to scan \ac{purl}s without the version fields.

\subparagraph{Expectation}
We expect that all \svstool{}s that support \ac{purl} will process \ac{purl} without the version field, and report the expected vulnerability, which in our case is \cve{CVE-2022-24434}. If the \svstool{} fails to report the expected vulnerability and does not explicitly inform the user of a potential error, we consider it a silent failure. Acceptable warning messages include \enquote{Error processing PURL for component X} or \enquote{Component Y: Invalid PURL}.

\medskip

\testcasetitle{21b5zfps}{Scenario 4}

\subparagraph{Summary}
\ac{SBOM} with two identifiers for a component, where vulnerable \ac{CPE} is before \ac{purl}.

\subparagraph{Description}
 This test case features a vulnerable \cpe{cpe:\ldots{}:\\8.11.4:...} and an unaffected \purl{pkg:\ldots{}@9.12.0}. The component \texttt{(maven) lucene-replicator}, describ-\\ed in this test case, is affected by \cve{CVE-2024-45772}.

\subparagraph{Rationale} 
The \href{https://cyclonedx.org/docs/1.6/json/}{CycloneDX schema} allows users to include two identifiers \ac{CPE} and \ac{purl} for a single component in an \ac{SBOM}. \ac{SBOM}-generating tools such as \href{https://github.com/anthonyharrison/distro2SBOM}{distro2SBOM} and \href{https://github.com/anchore/syft}{Syft} generate \ac{SBOM}s containing both \ac{CPE} and \ac{purl} identifiers for each component.

\subparagraph{Motivation}  
As some \svstool{}s, such as \dtrack{} and \grype{}, utilize both \ac{CPE} and \ac{purl} for vulnerability lookup, and since the CycloneDX schema allows users to include both identifiers, we aim to understand which identifier each \svstool{} prefers. If a \svstool{} prefers \ac{CPE}, it will report the expected vulnerability. However, if it prefers \ac{purl}, no vulnerability may be reported even if the component is actually vulnerable. Furthermore, based on the \href{https://github.com/DependencyTrack/dependency-track/issues/3445}{dependency-track\#3445} GitHub issue, where the order of sections in the \ac{SBOM} affects the result display, we will alter the order of the identifiers in the remaining test cases that fall under the same scenario (Scenario 4), in order to better analyze the behavior of each \svstool{}.

\subparagraph{Expectation}
In this test case, the \ac{CPE} appears before the \ac{purl}, with a vulnerable version \texttt{8.11.4}, while the \ac{purl} has a fixed version. We expect all \svstool{}s to report the expected vulnerability, which is \cve{CVE-2024-45772}.

\medskip

\testcasetitle{9xhb7rgj}{Scenario 4}

\subparagraph{Summary}
\ac{SBOM} with two identifiers for a component, where \ac{purl} is before vulnerable \ac{CPE}.

\subparagraph{Description}
Identical to the test case \testcase{21b5zfps}.

\subparagraph{Rationale} 
Identical to the test case \testcase{21b5zfps}.

\subparagraph{Motivation}  
Identical to the test case \testcase{21b5zfps}.

\subparagraph{Expectation}
In this test case, the \ac{purl} appears before the \ac{CPE}, where \ac{CPE} is with a vulnerable version \texttt{8.11.4}, while the \ac{purl} has a fixed version. We expect all \svstool{}s to report the expected vulnerability, which is \cve{CVE-2024-45772}.

\medskip

\testcasetitle{pq3cy9or}{Scenario 4}

\subparagraph{Summary}
\ac{SBOM} with two identifiers for a component, where \ac{CPE} is before vulnerable \ac{purl}.

\subparagraph{Description}
This test case features an affected \ac{purl} and a non-affected \ac{CPE}: \cpe{cpe: \ldots{} :9.12.0: \ldots{}} and \purl{pkg:\ldots{}@8.11.4}. The component
\componentp{maven}{lucene-\\replicator}, described in this test case, is affected by \cve{CVE-2024-45772}.

\subparagraph{Rationale} 
Identical to the test case \testcase{21b5zfps}.

\subparagraph{Motivation}  
The motivation for this test case is similar to that of the test case \testcase{21b5zfps}, with the distinction that if the \svstool{} utilizes \ac{CPE}, it fails to report the expected vulnerability, despite the component being vulnerable. In contrast, if the tool utilizes \ac{purl}, it successfully reports the expected vulnerability. This behavior is used to infer the identifier preference of the tool based on its vulnerability detection outcome.

\subparagraph{Expectation}
In this test case, the \ac{CPE} appears before the \ac{purl}, where \ac{purl} is with a vulnerable version \texttt{8.11.4}, while the \ac{CPE} has a fixed version. We expect all \svstool{}s to report the expected vulnerability, which is \cve{CVE-2024-45772}.

\medskip

\testcasetitle{5q46iw4f}{Scenario 4}

\subparagraph{Summary}
\ac{SBOM} with two identifiers for a component, where the vulnerable \ac{purl} is before \ac{CPE}.

\subparagraph{Description}
This test case features an affected \ac{purl} and a non-affected \ac{CPE}: \cpe{cpe: \ldots{} :9.12.0: \ldots{}} and \purl{pkg:\ldots{}@8.11.4}. The component
\componentp{maven}{lucene-\\replicator}, described in this test case, is affected by \cve{CVE-2024-45772}.

\subparagraph{Rationale} 
Identical to the test case \testcase{21b5zfps}.

\subparagraph{Motivation}  
Identical to the test case \testcase{pq3cy9or}.

\subparagraph{Expectation}
In this test case, the \ac{purl} appears before the \ac{CPE}, where \ac{purl} is with a vulnerable version \texttt{8.11.4}, while the \ac{CPE} has a fixed version. We expect all \svstool{}s to report the expected vulnerability, which is \cve{CVE-2024-45772}.

\medskip

\testcasetitle{sqs4tbob}{Scenario 5}

\subparagraph{Summary}
The \ac{SBOM} contains two components, both of which lack unique identifiers.

\subparagraph{Description}
This test case features two components, \texttt{(npm) dicer} and \componentp{npm}{multer}, both of which lack identifiers. The component described as \componentp{npm}{dicer} in this test case is affected by \cve{CVE-2022-24434}.

\subparagraph{Rationale} 
 \href{https://cyclonedx.org/docs/1.6/json/}{CycloneDX 1.6} does not require components to have an identifier of any kind (i.e., purl and cpe are both optional).

\subparagraph{Motivation}  
As most \ac{SBOM}-generating tools produce \ac{SBOM} with both \ac{CPE} and \ac{purl}, but we have observed instances where both identifiers were missing.
We have analyzed \ac{SBOM}s generated by a medium-sized security service provider (annual turnover \textgreater\;1~billion~Euro) confidentially shared with us and noted the absence of both \ac{CPE} and \ac{purl} entries in the \acp{SBOM}. 
Searching online, we found public evidence that this is not a unique experience.
For example, a \href{https://github.com/microsoft/vcpkg/issues/39254}{GitHub Issue\#39254} for Microsoft Vcpkg highlighted that Vcpkg-generated \ac{SPDX} \ac{SBOM}s did neither include a \ac{CPE} nor \ac{purl}, rendering the \ac{SBOM}s ineffective for \svs{}.
The JSON \ac{SBOM} shared in \href{https://github.com/e-m-b-a/emba/issues/734}{GitHub Issue\#734} related to EMBA's firmware security analyzer also contained component entries with only the CycloneDX \texttt{name} and \texttt{version} fields, lacking any component identifier. 

\subparagraph{Expectation}
This test case involves components that lack explicit \texttt{unique identifiers}. It is expected that an \svstool{} may generate these identifiers, since tools like \grype{} can produce \ac{CPE} entries when the \texttt{--add-cpes-if-none} flag is enabled. The \componentp{npm}{dicer} component includes all the necessary attributes (i.e., name, vendor, and version) required for \ac{CPE} generation. If the \svstool{} successfully reports the \cve{CVE-2022-24434} vulnerability, the test is considered passed, as this indicates correct behavior. Conversely, if the vulnerability is not reported, it is regarded as a silent failure.

\medskip

\testcasetitle{hawmnwbz}{Scenario 6}

\subparagraph{Summary}
The \ac{SBOM} contains a non canonical \ac{purl} string.

\subparagraph{Description}
This test case features a component, \texttt{lucene-replicator} with a valid but not canonical \ac{purl} string: \purl{pkg://\ldots}.\\ The described component is affected by \cve{CVE-2024-45772}.

\subparagraph{Rationale} 
The \href{https://github.com/package-url/purl-spec/blob/main/PURL-SPECIFICATION.rst}{purl specification} defines \purl{pkg:\ldots} as the standard scheme prefix. Nonetheless, the specification requires parsers to read non-canonical purl strings such as \purl{pkg://\ldots} ignoring and removing any extraneous slashes following the scheme. 

\subparagraph{Motivation}  
Non-canonical \ac{purl} strings, such as those with extra slashes, are allowed by the specification but the authors were unable to identify any non-canonical purl string in any of the \ac{BOM}s we analyzed. This may be due to \ac{BOM}s and purl strings being automatically generated by software tools and thus being less prone to human errors.

\subparagraph{Expectation}
Since non-canonical variants, such as \purl{pkg:// \ldots{}}, are permitted, we expect all \svstool{}s that utilize \ac{purl} to report the expected vulnerability, \cve{CVE-2024-45772}.

\medskip

\testcasetitle{qbqy99do}{Scenario 7}

\subparagraph{Summary}
The CycloneDX \ac{vex} with no influence.

\subparagraph{Description}
The \ac{vex} includes the component \texttt{(maven) lucene-replicator}, identified by both \ac{CPE} and \ac{purl}, with the vulnerable version \texttt{8.11.4}. The \texttt{affects} section of the \ac{vex} states that version \texttt{8.11.4} is affected, with the vulnerability \cve{CVE-2024-45772}.

\subparagraph{Rationale}
CycloneDX allows to embed \ac{vex} information in the \href{https://cyclonedx.org/docs/1.6/json/#vulnerabilities_items_affects}{vulnerabilities section} of the \ac{BOM}.
While few tools implement automated processing of CycloneDX \ac{vex} data, several \svstool{}s including \href{https://github.com/DependencyTrack/dependency-track}{Dependency-Track} and \href{https://docs.fossa.com/docs/generating-sboms#embedding-vdr--vex-statements}{\fossa{}}, can generate and export a CycloneDX \ac{vex} document.

\subparagraph{Motivation}  
There are \ac{vex} \ac{SBOM}s available in the CycloneDX \href{https://github.com/CycloneDX/bom-examples/tree/master/VEX}{BOM Examples} Github repository. The authors were not able to find any real-world \ac{BOM} with embedded CycloneDX \ac{vex} data.
Nevertheless, the concept of \ac{vex} is well established; for example large companies such as Cisco publish \ac{vex} documents, as noted in their \href{https://sec.cloudapps.cisco.com/security/center/resources/vex-cvr-faqs}{VEX FAQ page}.

\subparagraph{Expectation}
This test serves as a baseline to evaluate whether the \svstool{}s can identify a specific vulnerability, provided it is not suppressed by \ac{vex}. The vulnerability report indicates that one dependency is affected by \cve{CVE-2024-45772}. Since the report explicitly marks the application as vulnerable, we expect all \svstool{}s to report this vulnerability accordingly.

\medskip

\testcasetitle{0vo0efli}{Scenario 7}

\subparagraph{Summary}
The vulnerability is suppressed using the CycloneDX \ac{vex}.

\subparagraph{Description}
The \ac{vex} includes the component \texttt{(maven) lucene-replicator}, identified by both \ac{CPE} and \ac{purl}, with the vulnerable version \texttt{8.11.4}. The \texttt{affects} section of the \ac{vex} states that version \texttt{8.11.4} is unaffected, with the vulnerability \cve{CVE-2024-45772}.

\subparagraph{Rationale} 
Identical to the test case \testcase{qbqy99do}.

\subparagraph{Motivation}  
Identical to the test case \testcase{qbqy99do}.

\subparagraph{Expectation}
This test serves as a baseline to evaluate whether the \svstool{} correctly parses appended vulnerability information and suppresses the associated warning. The vulnerability report includes one dependency affected by \cve{CVE-2024-45772}; however, it explicitly states that the application is not vulnerable. Therefore, it is expected that all \svstool{}s suppress this vulnerability accordingly.

\medskip

\testcasetitle{omwcmwv1}{Scenario 8}

\subparagraph{Summary}
An \ac{SBOM} with out-of-order BOM segments.

\subparagraph{Description}
This \ac{SBOM} includes two components: \texttt{(npm) dicer} and \texttt{(npm) multer}. The \texttt{dicer} component is identified using both \ac{CPE} and \ac{purl}, and version \texttt{0.3.0} is affected by a known vulnerability, \cve{CVE-2022-24434}. The SBOM is structured in the following order: \texttt{metadata}, \texttt{vulnerabilities}, \texttt{dependencies}, and \texttt{components}. This order matches the actual layout of the SBOM used in this test case.

\subparagraph{Rationale} 
ECMA standard defines \texttt{CycloneDX v1.6} BOM \href{https://ecma-international.org/wp-content/uploads/ECMA-424_1st_edition_june_2024.pdf#page=16}{specifications}, where object types are arranged in order, however JSON (\href{https://www.rfc-editor.org/rfc/pdfrfc/rfc7159.txt.pdf#page=3}{RFC 7159}) does not guarantee order.

\subparagraph{Motivation}  
\ac{SBOM}-generating tools may produce \acp{SBOM} in accordance with the CycloneDX specifications as seen from CycloneDX \href{https://github.com/CycloneDX/bom-examples/tree/master/SBOM}{BOM-Examples} GitHub repository, maintaining the correct order of BOM segments. However, there have been instances where these segments appear out of order. For example the problem reported by a user in \href{https://github.com/DependencyTrack/dependency-track/issues/3445}{Dependency-Track \#3445} is caused by an \ac{SBOM} with out of order root elements. 
Similarly, we found multiple \ac{SBOM} with misplaced \texttt{metadata} sections (wrongfully) located at the end of the \ac{BOM}, in \ac{BOM}s obtained from a large European enterprise.

\subparagraph{Expectation}
We expect that an \svstool{} should reject malformed CycloneDX \acp{SBOM}, as the CycloneDX specification defines a required order for BOM segments. An appropriate error, such as \enquote{Invalid BOM segment} or \enquote{Error in uploading the BOM}, is expected in such cases. If such an error is observed, the \svstool{} is considered to have passed the test case. However, since JSON does not guarantee the order of elements, it is also possible that a vulnerability such as \cve{CVE-2022-24434} may still be encountered.

\medskip

\testcasetitle{3fvslnon}{Scenario 8}

\subparagraph{Summary}
An \ac{SBOM} with invalid root-level BOM segments.

\subparagraph{Description}
The description matches that of \testcase{omwcmv1}, except that the \texttt{licenses} and \texttt{properties} fields are not included within the \texttt{metadata} section of the BOM.

\subparagraph{Rationale} 
Identical to the test case \testcase{omwcmv1}.

\subparagraph{Motivation}  
This test case is identical to \testcase{omwcmv1}. However, such scenarios with invalid root-level BOM segments are rarely encountered in real-world \ac{SBOM}s. The primary motivation is to evaluate whether \svstool{} can detect that \ac{SBOM} is invalid. 

\subparagraph{Expectation}
Identical to the test case \testcase{omwcmv1}.

\subsection{Summary of Disclosures and Post-Disclosure Changes}
During the analysis of our test cases, we encountered several issues, one of which involved a valid \ac{CPE} not reporting the expected vulnerability, for instance, \cve{CVE-2022-24434} in \grype{}. Upon further investigation, we found that the expected vulnerability was present in the \href{https://github.com/anchore/grype-db}{Grype database}. This discrepancy prompted us to conduct a detailed investigation, which was documented through GitHub tickets. In response, we received feedback from their team regarding the issues, and they considered this a bug, which they subsequently fixed. All these interactions are recorded in \autoref{tab:disclosure}, where the reader can find three types as follows:

\begin{itemize}
    \item \textbf{Bug}  refers to an issue that we reported and the vendor acknowledged as a bug or implementation flaw.
    
    \item \textbf{Information} refers to details or observations that we shared, but which do not necessarily qualify as a bug. These include suggestions for improvement.
    
    \item \textbf{Related} refers to changes made by the \svstool{} developers after we disclosed our findings but without explicit confirmation that these are because of our research.
\end{itemize}

\begin{table*}[tbp]  
    \caption{Overview of the Practical Impact}
    \label{tab:disclosure}
    \centering
    \small
    \begin{tabular}{
        c  
        >{\RaggedRight}p{0.1\linewidth}  
        >{\RaggedRight}p{0.10\linewidth}  
        >{\RaggedRight}p{0.10\linewidth}  
        >{\RaggedRight}p{0.50\linewidth}  
    }
    \toprule
        \# & Reference & Test Case ID/SVS-Tool & Type & Description \\
    \midrule
        1 & {\href{https://github.com/anchore/grype/issues/2461}{grype:\#2461}} & u8h8dnoj & Bug & Tool incorrectly flagged a valid CPE as invalid due to inaccurate regex. Issue acknowledged, plan to switch to a more reliable parser. \\ 
        2 &  {\href{https://github.com/anchore/grype/issues/2434}{grype:\#2434}} & an7esfjj & Bug & In some conditions, tool incorrectly suppresses results from \ac{CPE} search. Fixed in version \version{0.88.0}. \\ 
        3 &  {\href{https://github.com/anchore/grype/pull/2591}{grype:PR\#2591}} & Grype & Related & The change allows empty versions to meet version constraints. Status of impact not confirmed. \\ 
        4 & --- & Vulert & Information & \svstool{} does not disclose its own version; the vendor acknowledged the feedback and confirmed plans to make it public. \\
        5 &  {\href{https://github.com/anchore/grype/issues/2462}{grype:\#2462}} & 0vo0efli & Information & Feature request: support for embedded VEX. No fix committed yet. \\ 
    \bottomrule
    \end{tabular}
\end{table*}





\subsection{Author Contributions}
Author contributions according to the Contributor Role Taxonomy (\href{https://credit.niso.org/}{CRediT}).
Martin Rosso: Conceptualization, Methodology, Investigation, Formal analysis, Writing (original draft / review \& editing).
Muhammad Asad Jahangir Jaffar: Investigation, Formal analysis, Data curation, Writing (original draft / review \& editing).
Alessandro Brighente: Conceptualization, Supervision, Writing (original draft / review \& editing).

\end{document}